\documentclass[11pt, twoside]{article}

\usepackage[english]{babel}
\usepackage[utf8]{inputenc}
\usepackage{natbib}
\usepackage{amsmath}
\usepackage{bbm}
\usepackage{graphicx}
\usepackage{multirow}
\usepackage{color}
\usepackage{xcolor}
\usepackage{svg}
\usepackage{natbib}
\usepackage{url}
\usepackage{float}
\usepackage{amsfonts}
\usepackage{booktabs}
\usepackage{listings}
\usepackage{lineno}
\usepackage[toc,page]{appendix}
\usepackage{enumitem}
\usepackage{amssymb}
\usepackage{pifont}
\usepackage[font=footnotesize]{subcaption}
\usepackage[font=footnotesize]{caption}
\usepackage{tikz}
\usepackage{authblk} 
\usepackage{appendix}
\usepackage[format=plain, labelfont={bf,it},
            textfont=it]{caption}
\usepackage{todonotes}
\usepackage{fancyhdr}

\newtheorem{theorem}{Theorem}

\newtheorem{algorithm}{Algorithm}
\newtheorem{remark}{Remark}
\newtheorem{example}{Example}

\usepackage[a4paper,margin=2.54cm]{geometry}

\begingroup\lccode`~=`_
\lowercase{\endgroup\def~}#1{_{\scriptscriptstyle#1}}

\AtBeginDocument{\mathcode`_="8000 \catcode`_=12 }

\usepackage{hyperref}
\hypersetup{
    colorlinks=true,
    linkcolor=blue,
    filecolor=red,
    citecolor=cyan,
    urlcolor=cyan,
}

\makeatletter
\def\Let@{\def\\{\notag\math@cr}}
\makeatother

\DeclareMathOperator*{\argmax}{arg\,max}

\newcommand{\ssum}[3][i]{\sum_{#1=#2}^{#3}}

\newcommand{\sign}[1]{\mathrm{sign}(#1)}

\newcommand{\row}[1]{\mathrm{row}(#1)}

\newcommand{\diag}[1]{\mathrm{diag}(#1)}

\newcommand{\bsy}[1]{\boldsymbol{#1}}
\newcommand{\mbf}[1]{\mathbf{#1}}
\newcommand{\mbb}[1]{\mathbb{#1}}
\newcommand{\mca}[1]{\mathcal{#1}}
\newcommand{\trm}[1]{\textrm{#1}}

\newcommand*{\medcap}{\mathbin{\scalebox{1.4}{\ensuremath{\cap}}}}%
\DeclareFontFamily{U}{mathx}{\hyphenchar\font45}
\DeclareFontShape{U}{mathx}{m}{n}{
      <5> <6> <7> <8> <9> <10>
      <10.95> <12> <14.4> <17.28> <20.74> <24.88>
      mathx10
      }{}
\DeclareSymbolFont{mathx}{U}{mathx}{m}{n}
\DeclareFontSubstitution{U}{mathx}{m}{n}
\DeclareMathAccent{\widecheck}{0}{mathx}{"71}
\DeclareMathAccent{\wideparen}{0}{mathx}{"75}

\title{Valid Post-Detection Inference for Change Points Identified Using Trend Filtering}

\author[1]{Reza Valiollahi Mehrizi \footnote{Corresponding Author: rvalioll@uwaterloo.ca}}
\author[1]{Shojaeddin Chenouri}
\affil[1]{Department of Statistics and Actuarial Science, University of Waterloo, Canada}

\newcommand\shorttitle{Valid Post-Detection Inference}
\newcommand\authors{Mehrizi and Chenouri}

\date{}

\begin{document}

\maketitle

\thispagestyle{empty}
\begin{abstract}
    There are many research works and methods about change point detection in the literature. However, there are only a few that provide inference for such change points after being estimated. This work mainly focuses on a statistical analysis of change points estimated by the PRUTF algorithm, which incorporates trend filtering to determine change points in piecewise polynomial signals. This paper develops a methodology to perform statistical inference, such as computing p-values and constructing confidence intervals in the newly developed post-selection inference framework. Our work concerns both cases of known and unknown error variance. As pointed out in the post-selection inference literature, the length of such confidence intervals are undesirably long. To resolve this shortcoming, we also provide two novel strategies, {\it global post-detection} and {\it local post-detection} which are based on the intrinsic properties of change points. We run our proposed methods on real as well as simulated data to evaluate their performances.  

\end{abstract}

\section{Introduction}

Change point detection has achieved considerable attention in recent years and found applications in many fields such as finance and econometrics, \cite{bai2003computation}, bioinformatics and genomics, \cite{futschik2014multiscale}, climatology, \cite{pezzatti2013fire}, and technology, \cite{galceran2017multipolicy}. This problem was first studied by \cite{page1954continuous,page1955test} and since then, has been of interest to many scientists including statisticians. 

Formally, change point detection seeks the locations of changes in the distribution of a signal for which an ordered sequence of observations is available. This paper considers change point detection in the mean of a distribution so that this underlying mean is a piecewise polynomial function. More specifically, let $\mbf y=(y_1,\, \ldots,\,y_n)$ be a response vector following the model
\begin{align}\label{fmodel.proj2}
    y_i=f_i+\varepsilon_i,\,\qquad\qquad i=1,\,\ldots,\,n\,,
\end{align}
where $f_i=f(i/n)$ is a deterministic and  unknown signal with equally spaced input points over the interval $[0,1]$. We assume that the error terms $\boldsymbol\varepsilon=(\varepsilon_1,\ldots,\varepsilon_n)$ are independent and have a Gaussian distribution with a mean of zero and finite variance $\sigma^2$ (either known or unknown).  We suppose that signal $f$ undergoes $J_0$ unknown and distinct changes at locations $0=\tau_0<\tau_1<\ldots<\tau_{J_0}<\tau_{J_0+1}=n$. Additionally, we allow the number of change points $J_0$ to grow with the sample size $n$. These change points divide the entire signal $f$ into $J_0+1$ sub-signals. We assume that each of these sub-signals are polynomial of order $r$ $(r=0,\,1,\,2,\,\ldots)$. We put no restriction over the signal at the locations of change points, i.e., the signal could be either continuous or discontinuous at change points. The focus in change point detection is on the estimation of the number and the positions of changes.

There is a vast and rich literature on change-point detection in statistical research. Depending on various aspects of the data at hand, many different techniques and algorithms exist in the literature. Greedy search techniques have long been applied to change point detection in various settings, in which change points are found one at a time. A traditional approach for the canonical change-point problem is the Binary Segmentation \cite{vostrikova1981detecting}. When the signal is piecewise constant, it is referred to as a canonical change-point problem in the literature. Recently refinement of Binary segmentation such as  Circular Binary Segmentation \cite{olshen2004circular}, and Wild Binary Segmentation \cite{fryzlewicz2014wild} have also been successfully applied.  More greedy search methodologies adapted for various setting have recently been introduced: \cite{li2016fdr}, \cite{fryzlewicz2018tail}, \cite{baranowski2019narrowest}, \cite{anastasiou2019detecting}, and \cite{fryzlewicz2018detecting}. 
Some authors have considered the canonical change-point detection as a penalized least square problem, examples are\cite{yao1989least}, \cite{davies2001local} \cite{harchaoui2010multiple} and \cite{qian2016stepwise}. 

Although there is a huge body of work on canonical change point framework, a more general case in which signal $f$ is modelled as a piecewise polynomial has attracted less attention. A few works, mainly focused on a linear piecewise model, exist in the literature; see \cite{bai1997estimating}, \cite{maidstone2017optimal}, \cite{baranowski2019narrowest} and  \cite{anastasiou2019detecting}. The recent work of \cite{mehrizi2020detection} introduces an approach called PRUTF, which is designed for change point detection for piecewise polynomial signals. PRUTF exploits the trend filtering problem introduced in \cite{tibshirani2014adaptive} and provides a novel algorithm to estimate change points. 

After analysts perform change point detection/estimation, genuine interest is to make inferences about the uncertainty of the detected/estimated points. Many research works such as \cite{eckley2011analysis}, \cite{horvath2014extensions}, and \cite{chen2001atomic} have discussed conducting such inferences. However, they have ignored the data-driven nature of the detected change points and regarded them as fixed locations. It is important to note that this data-dependent property of detected change points invalidates the inferences mentioned earlier. There is a new and rapidly growing framework that provides tools for conducting inference after a selection procedure. This type of inference is called {\it post-selection inference} and has been mainly developed for inference after variable selection in high-dimensional regression. See for example \cite{berk2013valid, lee2016exact, tibshirani2016exact, fithian2014optimal}. In this paper we adopt this body of work for inference after change point detection in piecewise polynomial signals. 

Interest in post-selection inference research ignited after the work of   \cite{potscher1991effects}. Later, in a sequence of articles, \cite{leeb2003finite,leeb2006can,leeb2008can} explored the estimation of the post-selection distribution. This subject has attracted much attention more recently as applications for model selection approaches have proliferated. \cite{berk2013valid} performed valid and conservative selective inference by considering all possible procedures that produce a selected model. \cite{lockhart2014significance} suggested an asymptotic test procedure to test whether a nonzero estimated coefficient in the lasso regression coincides with the true nonzero coefficient. \cite{lee2016exact} developed an exact test based on truncated Gaussian distribution for the solution path of the lasso at a fixed value of the regularization parameter $\lambda$. Concurrently, a similar test is proposed by \cite{tibshirani2016exact} for the lasso, LARs and forward stepwise regression with a fixed number of steps in the solution path. \cite{fithian2014optimal} studied the theoretical properties of post-selection inference and generalized the framework to the broad class of the exponential family of distributions. A number of extensions to different frameworks and applications are given in  \cite{loftus2014significance}, \cite{tian2017asymptotics}, \cite{tibshirani2018uniform}, \cite{hyun2018exact}, \cite{liu2018more}, \cite{bachoc2020uniformly}.

In a post-selection inference setting, since data have been used to fit a model, a conditional inference is required to restrict the sample space to unused data. This conditional approach prevents an inference procedure from using data twice (once for selection and once for inference). As a result of this conditioning, the post-detection distribution changes from Gaussian to truncated Gaussian \citep{lee2016exact, tibshirani2016exact}, reflecting the restriction on the sample space. However, it turns out that this truncation is often very severe and leads statistical tests to lose power and their respective confidence intervals to become undesirably wide; see Section \ref{sec:optimal.conditioning.proj2}. These problems occur because most of the data are employed to select the model. Consequently, there is insufficient information to use for drawing conclusions after the selection procedure. 

The application of post-selection inference in the  change points context, referred to as post-detection inference, was first addressed in \cite{hyun2018exact}. Their work, which is our main inspiration, applies the exact truncated Gaussian test for a given and selected number of steps in the solution path of generalized lasso. Similarly, \cite{hyun2018post} study post-detection inference for some popular change point detection methodologies such as Binary Segmentation, Circular Binary Segmentation, and Wild Binary Segmentation. 

In this paper, we study the problem of conducting valid statistical inferences for detected change points. For change point detection/estimation we use the {\it Pattern Recovery Using Trend Filtering} (PRUTF) algorithm introduced in \cite{mehrizi2020detection}. At the core of our framework, an important result states that the set of change points detected by PRUTF constitutes a polyhedron (a convex cone). We apply the post-selection inference framework to compute valid post-detection p-values as well as confidence intervals.

\subsection{Summary}
\label{sec:contribution.proj2}
We make the following contributions in this work.

\begin{itemize}
    \item  One fundamental aspect of our contribution is establishing that the set of change points identified using the PRUTF algorithm characterizes a polyhedron set in $\mbf y$. This characterization allows us to use the post-selection inference methodology for conducting statistical inference after change point detection. We emphasize that, to the best of our knowledge, the inference procedures for the significance of detected change points developed in this paper are the first of their kind for piecewise polynomial signals. \cite{mehrizi2020detection} also developed a stopping criterion for selecting the number of required steps for the PRUTF algorithm, which uses the Gaussian bridge property of dual variables. We will show that this criterion also forms a polyhedron and, consequently, applies to the post-detection inference framework.
    
    \item The implication of post-detection inference for the Gaussian model in \eqref{fmodel.proj2} leads us to propose two test statistics, one for known $\sigma^2$ and another for unknown $\sigma^2$. A significant feature of these test statistics is that their exact and finite sample distributions under the null hypothesis are $U(0,1)$. These test statistics allow us to conduct inference for the significance of change points estimated by PRUTF.
    
    \item We inspect the produced p-values and confidence intervals of detected change points using the two test statistics for both known and unknown noise variance cases. We show that the confidence intervals derived by these approaches are unacceptably wide. This also leads to a loss in the power of hypothesis tests. To resolve these shortcomings, we introduce two new methods of conditioning for post-detection inference. We call the first method {\it global post-detection} that focuses only on the target change point and makes inference regardless of the other change points. We call the second method {\it local post-detection} which takes into account detected locations that are the most relevant to the target change point.
    
    \item We conduct a comprehensive simulation study to investigate the performance of the proposed procedures in terms of power of post-detection tests and length of their confidence intervals. We also demonstrate applications of our algorithms to several real data sets.
    
\end{itemize}

The rest of this paper is structured as follows. We first provide a version of the PRUTF algorithm of \cite{mehrizi2020detection} along with a stopping criterion in Section \ref{sec:preliminary.proj2}. We also review conducting post-selection inference based on the polyhedron characterization of a selection procedure. In Section 3, we explain how to capture the representation of the corresponding polyhedron of the PRUTF algorithm. This representation also determines cases where the number of steps for the algorithm is adaptively selected using the stopping rule. Next, two different approaches are established in Section \ref{sec:post.detection.proj2} to conduct valid post-detection inference when the error variance is assumed both known and unknown. In Section \ref{sec:optimal.conditioning.proj2}, we discuss the shortcomings of the post-selection inference, conditional on the entire selected model and its signs, such as reduction in the power of tests and wide confidence intervals. We then propose two strategies to resolve these shortcomings. Section \ref{sec:simulations} provides some numerical investigation, including real data analysis and a simulation study. We conclude the work with a discussion in Section \ref{sec:discussion}.

\section{Notations and Preliminaries}
\label{sec:preliminary.proj2}

\subsection{Notations}
\label{sec:notation.proj2}
The following notations are used throughout the paper. For an $m\times n$ matrix $\mbf A$, a sub-matrix $\mbf A_{\mca I}$ extracts the rows of the matrix $\mbf A$ that are in the set  $\mca I \subseteq\{1,\ldots,m\}$. Similarly, a sub-vector $\bsy a_{\mca I}$ picks the components of a vector $\bsy a$ that are in $\mca I$. Quantities $\mbf A_{-\mca I}$ and $\bsy a_{-\mca I}$ select rows and elements of $\mbf A$ and $\bsy a$ that are not in $\mca I$, respectively. Additionally, for selecting $i$-th row of $\mbf A$, the notation $[\mbf A]_i$ and for its  $(i,j)$-th element the notation $[\mbf A]_{ij}$ are used. Moreover, for a real number $x$, $\lfloor x\rfloor$ denotes the greatest integer less than or equal to $x$. Notation $\lceil x \rceil$ also gives the smallest integer greater or equal to $x$.

\subsection{Solution Path of Trend Filtering}
\label{sec:solutionpath.proj2}
In this section, we briefly review the methodology introduced in \cite{mehrizi2020detection}. Suppose a response vector $\mbf y$ and a penalty matrix $\mbf D^{(r+1)}$, to be defined, are given. The PRUTF estimate of change points in a piecewise polynomial signal $\mbf f$ of order $r$ is nonzero entries of $\mbf D^{(r+1)} \, \widehat{\mbf f}$, where $ \widehat{\mbf f}$ is the the solution of the the following penalized least square problem 
\begin{align}\label{tf.obj.proj2}
    \min_{\mbf f\in \mathbb{R}^n}\,
    \frac{1}{2}\, \big\| \, \mathbf{y}\, -\, \mbf f \, \big\|_2^2+ \lambda\, \big\| \, \mathbf{D}^{(r+1)}\, \mbf f \, \big\|_1\,.
\end{align}
In this optimization problem known as {\it trend filtering}, $\lambda$ is the regularization parameter used to control the balance between sparsity and model fit, and $(n-r-1)\times n$ matrix $\mbf D^{(r+1)}$ is the $(r+1)$-th discrete difference operator. The first order discrete difference operator matrix $\mathbf{D}^{(1)}$ is defined as 
\begin{align*}
    \mathbf{D}^{(1)}=\begin{pmatrix}
    -1 & 1 & 0 & \ldots & 0 & 0 \\
    0 & -1 & 1 & \ldots & 0 & 0 \\
    \vdots & & & & & \vdots \\
    0 & 0 & 0 & \ldots & -1 & 1 \\
    \end{pmatrix}.
\end{align*}
For $r\geq 1$, the difference operator of order $r+1$ can be recursively computed by $\mathbf{D}^{(r+1)}=\mathbf{D}^{(1)}\times \mathbf{D}^{(r)}$, where $\mathbf{D}^{(1)}$ here is the $(n-r-2)\times (n-r-1)$ version of the first difference operator. Henceforth, for simplicity, we suppress the superscript for $\mathbf{D}^{(r+1)}$, i.e., $\mbf D=\mathbf{D}^{(r+1)}$

As a particular case of trend filtering ($r=0$), one-dimensional fused lasso has been effectively used to detect change points and produce a piecewise constant estimate of a signal \citep{harchaoui2010multiple, rinaldo2009properties, qian2016stepwise}. Recently, \cite{mehrizi2020detection} applied trend filtering to detect change points in a piecewise polynomial signal. The method, called PRUTF, returns estimates of change points along the trend filtering solution path.

Problem \eqref{tf.obj.proj2} has a unique solution because $\mbf D$ is a full row rank matrix; see \cite{ali2019generalized} for more details. One way to estimate the signal $\mbf f$ is to solve its equivalent Lagrange dual problem. Following the argument of \cite{tibshirani2011solution} and duality conditions of \cite{boyd2004convex}, the dual problem of trend filtering can be written as 
\begin{align}\label{tf.dual.obj.proj2}
    \min\limits_{\mathbf{u}\in \mathbb{R}^{m}} \, \frac{1}{2}\, \big\| \, \mathbf{y}\, -\, \mathbf{D}^T\mathbf{u} \, \big\|_2^2   \qquad\qquad \textrm{subject to } \qquad\qquad \big\| \, \mbf u \, \big\|_\infty\leq\lambda\,,
\end{align}
which is a quadratic problem constrained within a box with side length $2\lambda$. The connections between the primal and dual solutions are 
\begin{align}\label{primal.dual.proj2}
    \widehat{\mbf u}_{\lambda}=\lambda \widehat{\bsy\gamma} 
    \qquad\qquad\qquad \trm{and} \qquad\qquad\qquad
    \widehat{\mbf f }_\lambda=\mathbf{y}-\mathbf{D}^T\widehat{\mathbf{u}}_\lambda\,,
\end{align}

where $\widehat{\bsy\gamma} \in \mathbb{R}^{m}$ is a subgradient of $\big\| \mbf x \big\|_1$ computed at $\mbf x=\mbf D \, \widehat{\mbf f }_\lambda$, given by
\begin{align}\label{gamma.subgrad.proj2}
    \widehat{\gamma}_i\in \left\{
    \begin{array}{lcl}
        \{+1\} & \textrm{if} & [\mathbf{D\widehat{\mbf f }_\lambda}]_i>0  \\
        \{-1\} & \textrm{if} & [\mathbf{D\widehat{\mbf f }_\lambda}]_i<0  \\
         \,[-1,+1] & \textrm{if} &  [\mathbf{D\widehat{\mbf f }_\lambda}]_i=0\,. 
    \end{array}
    \right.
\end{align}
\cite{tibshirani2011solution} proposed an algorithm to compute the entire dual solution path along the regularization parameter $\lambda$.  PRUTF modifies this algorithm by removing $r+1$ rows of $\mbf D$ once a coordinate joins the boundary set. 
This is in contrast with the original algorithm of \cite{tibshirani2011solution} in which only one row of $\mbf D$ is removed. The modification in PRUTF makes the method well-suited for the discovery of change points; see \cite[Remark 3]{mehrizi2020detection}. Below, we briefly review the PRUTF algorithm to identify change points along the solution path as the value of $\lambda$ descends. For more details, see \cite{mehrizi2020detection}.

\begin{algorithm}\label{alg:tf.dual.path}
{\bf (Dual Solution Path of Trend Filtering)}

Define $r_b=\lceil (r+1)/2\rceil-1$ and $r_a=\lfloor (r+1)/2\rfloor$, and also set $\widehat{\bsy\tau}_0= \emptyset$, the empty set. 
\begin{enumerate}
    \item At step $j=1$, initialize the boundary set $\mathcal{B}_1=\{\widehat{\tau}_{1}-r_b,\,\widehat{\tau}_{1}-r_b+1,\,
    \ldots,\widehat{\tau}_1\}$ and its associated sign vector $\widehat{\mbf s}_{\mathcal{B}_1}=\{\widehat{s}_1,\ldots,\widehat{s}_1\}$, both with cardinality of $r_b+1$, where $\widehat{\tau}_1$ is obtained by
    \begin{align*}
        \widehat{\tau}_1=\argmax_{\,i=1,\,\ldots,\,m} \big|\, \widehat{u}_i\, \big| \,,
    \end{align*}
    and $\widehat{s}_1=\sign{ \widehat{u}_{_{ \widehat{\tau}_1}}}$, where $\widehat{u}_i$ is the $i$-th element of the vector $\widehat{\mathbf{u}}=\left(\mathbf{DD}^T\right)^{-1}\mathbf{D}\,\mathbf{y}$. In order to obtain the locations of change points, we maintain the set $\widehat{\bsy\tau}_1=\{\widehat{\tau}_1\}$. We also record the first joining time $\lambda_1=|\widehat{u}_{\widehat{\tau}_1}|$ and keep track of the augmented boundary set $\mathcal{A}_1=\{\widehat{\tau}_1-r_b,\ldots,\widehat{\tau}_1+r_a\}$ and its corresponding sign vector $\widehat{\mbf s}_{\mathcal{A}_1}= \{\widehat{s}_1, \ldots,\widehat{s}_1\}$ of length $r+1$. 
    
    \item For step $j=2,\,3,\,\ldots\,,$
    \begin{enumerate}
    \item Obtain the pair $(\widehat{\tau}_{j}^{\,\mathrm{join}},\widehat{s}_{j}^{\,\mathrm{join}})$ from
    \begin{align}\label{joinpair.proj2}
        (\widehat{\tau}_{j}^{\,\mathrm{join}},\widehat{s}_{j}^{\,\mathrm{join}})=\argmax_{\,i\notin \mathcal{A}_{j-1},~s\in\{-1,1\}}~ \frac{a_i}{s+b_i}\cdot\mathbbm{1}\left\{0\leq \frac{a_i}{s+b_i} \leq \lambda_{j-1}\right\},
    \end{align}
    and set the next joining time $\lambda_{j}^{\,\mathrm{join}}$ as the value of $\frac{a_i}{s+b_i}$, for $i=\widehat{\tau}_{j}^{\,\mathrm{join}}$ and $s= \widehat{s}_{j}^{\,\mathrm{join}}$. The quantities $\bsy a$ and $\mbf b$ in the above equation are defined as 
    \begin{align}\label{ab.pro2}
        \bsy a= \big( \mathbf{D}_{-\mca A}\, \mathbf{D}_{-\mca A}^T \big)^{-1}\mathbf{D}_{-\mca A}\,\mathbf{y}\,
        \qquad \trm{and} \qquad
        \mathbf{b}= \left(\mathbf{D}_{-\mca A}\, \mathbf{D}_{-\mca A}^T\right)^{-1} \mathbf{D}_{-\mca A}\, \mathbf{D}_{\mca A}^T\, \mathbf{s}_{\mca A}\,.
    \end{align}
    
    \item Obtain the pair $(\widehat{\tau}_{j}^{\,\mathrm{leave}},\widehat{s}_{j}^{\,\mathrm{leave}})$ from 
    \begin{align}\label{leavepair.proj2}
        (\widehat{\tau}_{j}^{\,\mathrm{leave}},\widehat{s}_{j}^{\,\mathrm{leave}})=\argmax_{i\in \mathcal{B}_{j-1},~s\in\{-1,1\}}~ \frac{c_i}{d_i}\cdot\,\mathbbm{1}\left\{c_i\leq 0~,~ d_i< 0\right\},
    \end{align}
    and assign  the next leaving time $\lambda_{j}^{\,\mathrm{leave}}$ as the value of $\frac{c_i}{d_i}$, for $i=\widehat{\tau}_{j}^{\,\mathrm{leave}}$ and $s=\widehat{s}_{j}^{\,\mathrm{leave}}$, where $\mbf c$ and $\mbf d$ are given by
    \begin{align}\label{cd.proj2}
        \mathbf{c}= \mathrm{diag}\, (\mathbf{s}_{\mathcal{B}})\mathbf{D}_{\mathcal{B}}\left(\mathbf{y}-\mathbf{D}_{-\mathcal{A}}^T\,\bsy{a}\right)
        \quad \trm{and} \quad
        \mathbf{d}= \mathrm{diag}\, (\mathbf{s}_{\mathcal{B}})\mathbf{D}_{\mathcal{B}}\left(\mathbf{D}_{\mathcal{A}}^T\,\mathbf{s}_{\mathcal{A}}-\mathbf{D}_{-\mathcal{A}}^T\,\mathbf{b}\right).
    \end{align}
    
    \item Let $\lambda_{j}=\max\{\lambda_{j}^{\,\mathrm{join}},\lambda_{j}^{\,\,\mathrm{leave}}\}$,
    then the boundary set $\mathcal{B}_{j}$ and its sign vector $\widehat{\mbf s}_{\mathcal{B}_{j}}$ are updated in the following fashion:
\begin{itemize}
\item[-- ] Either append $\{\widehat{\tau}_{j}^{\,\mathrm{join}}-r_b,\,\widehat{\tau}_{j}^{\,\mathrm{join}}-r_b+1,\,\ldots,\widehat{\tau}_{j}^{\,\mathrm{join}}\}$ and the corresponding signs $\{\widehat{s}_{j}^{\,\mathrm{join}},\ldots,\widehat{s}_{j}^{\,\mathrm{join}}\}$ to $\mathcal{B}_{j-1}$ and $\widehat{\mbf s}_{\mathcal{B}_{j-1}}$, respectively, provided that $\lambda_{j}=\lambda_{j}^{\mathrm{join}}$. Also, add $\widehat{\tau}_{j}^{\mathrm{join}}$ to $\widehat{\bsy\tau}_{j-1}$. 
\item[-- ] Or remove $\{\widehat{\tau}_{j}^{\,\mathrm{leave}},\, \widehat{\tau}_{j}^{\,\mathrm{leave}}+1
    \ldots ,\widehat{\tau}_{j}^{\,\mathrm{leave}}+r_b\}$ and the corresponding signs $\{\widehat{s}_{j}^{\,\mathrm{leave}}, \linebreak, \ldots,
    \widehat{s}_{j}^{\,\mathrm{leave}}\}$ from $\mathcal{B}_{j-1}$ and $\widehat{\mbf s}_{\mathcal{B}_{j-1}}$, respectively, provided that $\lambda_{j}=\lambda_{j}^{\,\mathrm{leave}}$. Also, eliminate $\widehat{\tau}_{j}^{\,\mathrm{leave}}$ from $\widehat{\bsy\tau}_{j-1}$.
\end{itemize}
In the same manner, the augmented boundary set, $\mathcal{A}_{j}$ and its sign, $\widehat{\mbf s}_{\mathcal{A}_{j}}$ are formed by adding $\left\{\widehat{\tau}_{j}^{\,\mathrm{join}}-r_b,\ldots, \widehat{\tau}_{j}^{\,\mathrm{join}}+r_a\right\}$  and $\left\{\widehat{s}_{j}^{\,\mathrm{join}},\ldots,\widehat{s}_{j}^{\,\mathrm{join}}\right\}$ to $\mathcal{A}_{j-1}$ and $\widehat{\mbf s}_{\mathcal{A}_{j-1}}$ if $\lambda_{j}=\lambda_{j}^{\,\,\mathrm{leave}}$ or, otherwise, by removing associated $\left\{\widehat{\tau}_{j}^{\,\mathrm{leave}},\ldots, \widehat{\tau}_{j}^{\,\mathrm{leave}}+ r\right\}$  and $\big\{\widehat{s}_{j}^{\,\mathrm{leave}},   \ldots $, $\widehat{s}_{j}^{\,\mathrm{leave}}\big\}$ from $\mathcal{A}_{j-1}$ and $\widehat{\mbf s}_{\mathcal{A}_{j-1}}$. The dual solution is computed as $\widehat{\mbf u}_{\mca A_j}(\lambda)=\bsy a-\lambda \mbf b$ for interior coordinates and $\widehat{\mbf u}_{-\mca A_j}(\lambda)=\lambda\, \widehat{\mbf s}_{\mca A_j}$ for boundary coordinates over $\lambda_j\leq\lambda\leq\lambda_{j-1}$. 
\end{enumerate}
\item Repeat Step 2 until $\lambda_j> 0$.
\end{enumerate}
\end{algorithm}
Along the dual solution path, the above algorithm keeps track of four sets: the set of dual change points denoted by $\widehat{\bsy\tau}$, the set of change points signs denoted by $\widehat{\mbf s}$, the set of coordinates that hits the boundary $\pm \lambda$, denoted by $\mca B$ and the set of boundary coordinates plus $r_a$ coordinates right after each change point, denoted by $\mca A$. The critical points $\lambda_1\geq\lambda_2\geq \ldots\geq 0$ are called knots and indicate the values of the regularization parameter at which the boundary set $\mca B$ changes. Notice that $\widehat{\bsy\tau}$ derived along the above algorithm delivers locations of change points for the dual problem of trend filtering. In order to obtain locations of change points in the underlying signal (primal trend filtering) which is our main interest, we require to add $r_b$ to any element of $\widehat{\bsy\tau}$, that is, $\{\widehat{\tau}_1+r_a, \widehat{\tau}_2+r_a, \ldots\}$. 


\subsection{Stopping Criterion}
\label{sec:stopping.proj2}
To prevent false discovery of change points by the PRUTF algorithm, we need a suitable criterion to end the search algorithm. \cite{mehrizi2020detection} designed a stopping rule based on the properties of dual variables within each block created by detected change points. More precisely, the authors have proven that the dual variables inside each block constitute a Gaussian bridge process. For more details, see Thoerem 1 in \cite{mehrizi2020detection}. It turns out that each entry of $\widehat{\mbf u}_{-\mca A}^{\,\trm{st}}= \left(\mathbf{D}_{-\mca A}\mathbf{D}_{-\mca A}^T\right)^{-1}\mathbf{D}_{-\mca A}\,\mathbf{y}$ tends to stay close to the zero line, after locating all change points. Therefore, at each step of Algorithm \ref{alg:tf.dual.path}, it is checked whether the maximum of $\left|\widehat{\mbf u}_{-\mca A}^{\,\trm{st}}\right|$ goes beyond a certain threshold. When there is no value to exceed this threshold, the algorithm will stop. More precisely, for $| \mca A_{_{j_0}} |$ as the cardinality of $\mca A_{_{j_0}}$ and $k=m-| \mca A_{_{j_0}} |$, the PRUTF algorithm terminates at the iteration $j_{_0}$ if
\begin{align*}
    \max_{0\leq t\leq 1} ~ \Big| \, \widehat{\mbf u}_{_{-\mca A_{j_{_0}}}}^{\,\trm{st}} \big( \lfloor\, kt\, \rfloor \big) \Big|\, \leq \, \sigma\, x_{_\alpha} \big( k-r \big)^{(2r+1)/2}\,, 
\end{align*}
where $x_{_\alpha}$ is derived from the equation
\begin{align}\label{thresh.stop.rule}
    \sum\limits_{i=1}^{\infty}(-1)^{i+1}\exp\left(\frac{-2\,i^2\,x_{_\alpha}^2}{S_r^2(k)}\right)=\frac{\alpha}{2}\,.
\end{align}
In the above equation, the quantity $S_r^2(k)$ is the $k$-th diagonal element of the matrix $\left(\mathbf{D}_{-\mca A}\mathbf{D}_{-\mca A}^T\right)^{-1}$.

\begin{remark}
\cite{mehrizi2020detection} have established that the PRUTF algorithm is not consistent in pattern recovery when the true signal $\mbf f$ contains staircase patterns. A staircase pattern refers to the phenomenon that the signs of two adjacent change points are in the same direction. This pattern causes PRUTF to identify many false change discoveries within segments with a staircase. Instead, they have proposed a modified algorithm called mPRUTF, which consistently recovers the true signal even in the presence of staircases.
\end{remark}

\subsection{Post-Selection Inference With Polyhedron Selection Procedures}
\label{polyhedron.selection.procedure}
In this section, we review some key concepts of post-selection
inference in a linear regression setting. For details about a broader class of models, see \cite{fithian2014optimal}. Suppose that the observations $\mbf y\in \mbb R^n$ follow a Gaussian regression model with either known or unknown variance. Also, let $\mca M$ be a finite collection of all possible models $\mbf M$ obtained from a model selection procedure, that is, $\mca M= \big\{ \mbf M:\, \mbf M\subseteq \{1,\ldots, p\} \big\}$, for $p$ to be the number of variables. The goal is, therefore, to carry out statistical inference for a selected model $\widehat{\mbf M} (\mbf y)=\mbf M$. Since we adaptively choose $\widehat{\mbf M} (\mbf y)$ using the data, for inference, it is natural to consider the conditional distribution given $\widehat{\mbf M} (\mbf y)$. This conditional distribution is called post-selection distribution. For testing null hypothesis $H_0$, we seek to control the post-selection type-I error, defined as $P_{H_0} \big( \trm{Reject } H_0\, \big |\, \widehat{\mbf M} (\mbf y)\in \mca M \big)$, at the nominal level $\alpha$. By analogy to the classical theory, a post-selection confidence interval can then be built by inverting the associated post-selection hypothesis test.

Throughout this paper, we restrict our attention to a specific class of selection procedures known as {\it polyhedron (affine) selection procedures}.
Again, suppose a variable selection approach picks a model $\widehat{\mbf M} (\mbf y)$ from a finite collection of models $\mca M$ using the data $\mbf y$. This selection approach is called a polyhedron selection procedure if any $\widehat{\mbf M} (\mbf y)\in \mca M$ can be characterized as a polyhedron set in respect to $\mbf y$. In other words, any selected model $ \widehat{\mbf M} (\mbf y)$ chosen by the polyhedron selection procedure can be written in the form of
\begin{align}\label{polyhed}
    \big\{\mbf y:~ \mbf A \mbf y\geq \mbf q \big\},
\end{align}
where the matrix $\mbf A \in \mbb R^{p\times n}$ and the vector $\mbf q\in \mbb R^{p}$ are dependent on $\mbf y$. Observe that the inequality in \eqref{polyhed} is interpreted componentwise. Selection approaches using the $\ell_1$-penalized generalized linear model, including Lasso \cite{lee2016exact, tibshirani2016exact}, LARS \cite{taylor2018post} and generalized lasso \cite{hyun2018exact},  are examples of such polyhedron selection procedures.

\section{Construction of Polyhedron}
\label{sec:polyhedron.proj2}
In this section, we describe how the change points detection procedure using PRUTF can be represented as a polyhedron. This representation enables us to state our post-detection inference conditional on a polyhedron. Suppose $\big\{ \widehat{\bsy\tau}_{j}\, ,\, \widehat{\mbf s}_j \big\}$ is the set of detected change points and their signs at step $j$ of the PRUTF algorithm. Below, we describe that 
$\big\{ \widehat{\bsy\tau}_{j}= \widehat{\bsy\tau}_{j}(\mbf y)\, ,\, \widehat{\mbf s}_j= \widehat{\mbf s}_j(\mbf y) \big\}$ as a function of $\mbf y$ is indeed a polyhedron. Thus, formally, for a matrix $\mbf A$ and vector $\mbf q$,
\begin{align}\label{polyhedron.proj2}
    \Big\{\mbf y:\, \widehat{\bsy\tau}_{j}= \widehat{\bsy\tau}_{j}(\mbf y)\, ,\, \widehat{\mbf s}_j= \widehat{\mbf s}_j(\mbf y) \Big\}=\Big\{\mbf y:\, \mbf A \mbf y\geq \mbf q \Big\}.
\end{align}
From now on, we call $\mbf A$ the polyhedron matrix. Observe that, in general, $\widehat{\bsy\tau}_j$ does not necessarily contain $j$ entries since (except in the case of $r=0$) the dual solution path of PRUTF can either add a change point to or remove it from the augmented boundary set at each step \citep{mehrizi2020detection}.

\subsection{Construction of Polyhedron Along the Solution Path}
\label{sec:polyhedron.solution.path}
In the following, the construction process of the matrix $\mbf A$ and vector $\mbf q$ associated with the polyhedron along the dual solution path of PRUTF is presented. We present this construction in steps according to the steps of Algorithm \ref{alg:tf.dual.path}.

\begin{enumerate}
    \item For $j=1$, the conditions for deriving the pair $(\widehat{\tau}_1\,,\, \widehat{s}_1)$ can be rewritten as
    \begin{align}\label{Gamma.step1.proj2}
        & \widehat{s}_1\left[\left(\mathbf{DD}^T\right)^{-1}\mathbf{D}\right]_{\widehat{\tau}_1}\mbf y ~ \geq ~ \pm \left[\left(\mathbf{DD}^T\right)^{-1}\mathbf{D}\right]_i\mbf y,
    \end{align}
    for any $i\neq \widehat{\tau}_1$.
    This implies that the polyhedron matrix after the first step, denoted by $\mbf A_1$, has $2(m-1)$ rows, formed by $\widehat{s}_1\left[\left(\mathbf{DD}^T\right)^{-1}\mathbf{D}\right]_{\widehat{\tau}_1}\pm \left[\left(\mathbf{DD}^T\right)^{-1}\mathbf{D}\right]_i\, $, for any $i\neq \widehat{\tau}_1$.
    
    \item For $j=1\, ,\, 2,\, \ldots \,$, assume that $\mbf A_{j}$ is the polyhedron matrix associated with $\big\{ \widehat{\bsy\tau}_{j}\, ,\, \widehat{\mbf s}_j \big\}$ at step $j$ of Algorithm \ref{alg:tf.dual.path}. In order to construct the polyhedron matrix for the set  $\big\{ \widehat{\bsy\tau}_{j+1}\, ,\, \widehat{\mbf s}_{j+1} \big\}$, a number of rows will be appended to $\mbf A_{j}$ according to Step 2 of Algorithm \ref{alg:tf.dual.path}. Recall that part (a) of Step 2 corresponds to the specification of the joining coordinate and its sign. Therefore, specifying $\widehat{\bsy\tau}_{j+1}$ is equivalent to satisfying the conditions
    {\footnotesize
    \begin{align}\label{Gamma.step2a.proj2}
        &\sign{a_i}\left[\left(\mbf D_{-\mca{A}_j}\mbf D_{-\mca{A}_j}^T\right)^{-1}\mbf D_{-\mca{A}_j}\right]_{i}\mbf y ~ \geq ~ 0,
        \qquad\qquad i\notin \mca A_j,
        \\[11pt]
        & \frac{\left[\left(\mbf D_{-\mca{A}_j}\mbf D_{-\mca{A}_j}^T\right)^{-1}\mbf D_{-\mca{A}_j}\right]_{\widehat{\tau}_{j+1}^{\,\textrm{join}}}\, \mbf y}{\widehat{s}_{j+1}^{\,\textrm{join}}+\left[\left(\mbf D_{-\mca{A}_j}\mbf D_{-\mca{A}_j}^T\right)^{-1}\mbf D_{-\mca{A}_j}\right]_{\widehat{\tau}_{j+1}^{\,\textrm{join}}}\mbf D_{\mca{A}_j}^T \widehat{\mbf s}_{_{\mca{A}_j}}}
        ~ \geq ~ 
        \frac{\left[\left(\mbf D_{-\mca{A}_j}\mbf D_{-\mca{A}_j}^T\right)^{-1}\mbf D_{-\mca{A}_j}\right]_{i}\mbf y}{\sign{a_i}+\left[\left(\mbf D_{-\mca{A}_j}\mbf D_{-\mca{A}_j}^T\right)^{-1}\mbf D_{-\mca{A}_j}\right]_{i}\mbf D_{\mca{A}_j}^T \widehat{\mbf s}_{_{\mca{A}_j}}},
    \end{align}}
    for any $i\notin \mca A_{j}\, \bigcup \, \{ \, \widehat{\tau}_{j+1}^{\,\textrm{join}} \,\}$. The vector $\bsy a$ in the aforementioned equations is given in \eqref{ab.pro2}.  The above inequalities will add $2\, \big( m-|\mca A_j| \big)-1$ rows to the matrix $\mbf A_{j}$.
    
    Part (b) of Step 2 explains conditions for the determination of the leaving coordinate. First, rows corresponding to conditions $c_i<0$ and $d_i<0$ for $i\in \mca B_{j}$ are added to $\mbf A_{j}$.
    To this end, we disregard those $i\in \mca B_{j}$ which $d_i\geq 0$ and partition the remaining entries into two groups. The first group called viable leaving coordinates includes those which $c_i<0$ and the second group as its complement. Denoting $\mbf L_{j+1}$ to be the collection of viable leaving coordinates, we consider
    \begin{align}
        \mbf L_{j+1}= \Big\{ \, i\in \mca B_j:~ d_i<0 \quad \textrm{and} \quad c_i<0 \, \Big\},
    \end{align}
    with the complementary set
    \begin{align}
        \mbf L_{j+1}^c= \Big\{ \, i\in \mca B_j:~ d_i<0 \quad \textrm{and} \quad c_i\geq 0 \, \Big\}.
    \end{align}
    
    Since vector $\mbf d$, defined in \eqref{cd.proj2}, does not depend on $\mbf y$, the conditions $c_i<0$ and $d_i<0$ can be expressed as
    {\footnotesize
    \begin{align}\label{gam.leaving.cond1}
        &\left[\diag{\widehat{\mbf s}_{\mathcal{B}_j}}\,\mathbf{D}_{\mathcal{B}_j}\left(\mathbf{I}-\mathbf{D}_{-\mathcal{A}_j}^T\left(\mbf D_{-\mca{A}_j}\mbf D_{-\mca{A}_j}^T\right)^{-1}\mbf D_{-\mca{A}_j}\right)\right]_{i} \mbf y ~ < ~ 0 \, ,\qquad \textrm{for}\quad i\in \mbf L_{j+1},
        \\[11pt]
        &\left[\diag{\widehat{\mbf s}_{\mathcal{B}_j}}\,\mathbf{D}_{\mathcal{B}_j}\left(\mathbf{I}-\mathbf{D}_{-\mathcal{A}_j}^T\left(\mbf D_{-\mca{A}_j}\mbf D_{-\mca{A}_j}^T\right)^{-1}\mbf D_{-\mca{A}_j}\right)\right]_{i}\mbf y ~ \geq ~ 0\, ,\qquad \textrm{for} \quad i\in \mbf L_{j+1}^c.
    \end{align}}
    The above equations correspond to $|\mbf L_{j+1}|+|\mbf L_{j+1}^c|$ rows of $\mbf A_{j+1}$. Second, the condition associated with the specification of the pair $\big( \widehat{\tau}_{j+1}^{\,\, \mathrm{leave}} \, ,\, \widehat{s}_{j+1}^{\,\, \mathrm{leave}} \big)$ must be added to the rows of $\mbf A_{j}$. This condition can be captured for any $i\in \mbf L_{j+1}\, \backslash\, \{\widehat{\tau}_{j+1}^{\,\,\mathrm{leave}}\}$ by the inequality
    {\footnotesize
    \begin{align}\label{gam.leaving.cond2}
        &\hspace{-2cm}\frac{\left[\diag{\widehat{\mbf s}_{\mathcal{B}_j}}\mathbf{D}_{\mathcal{B}_j}\left(\mathbf{I}-\mathbf{D}_{-\mathcal{A}_j}^T\left(\mbf D_{-\mca{A}_j}\mbf D_{-\mca{A}_j}^T\right)^{-1}\mbf D_{-\mca{A}_j}\right)\right]_{\widehat{\tau}_{j+1}^{\,\textrm{leave}}}\mbf y}{\left[\diag{\widehat{\mbf s}_{\mathcal{B}_j}}\mathbf{D}_{\mathcal{B}_j}\left(\mathbf{I}-\mathbf{D}_{-\mathcal{A}_j}^T\left(\mbf D_{-\mca{A}_j}\mbf D_{-\mca{A}_j}^T\right)^{-1}\mbf D_{-\mca{A}_j}\right)\right]_{\widehat{\tau}_{j+1}^{\,\textrm{leave}}}\mbf D_{\mca{A}_j}^T \widehat{\mbf s}_{\mca{A}_j}} ~ \geq ~ 
        \\[10pt]
        &\qquad\qquad\qquad\frac{\left[\diag{\widehat{\mbf s}_{\mathcal{B}_j}}\mathbf{D}_{\mathcal{B}_j}\left(\mathbf{I}-\mathbf{D}_{-\mathcal{A}_j}^T\left(\mbf D_{-\mca{A}_j}\mbf D_{-\mca{A}_j}^T\right)^{-1}\mbf D_{-\mca{A}_j}\right)\right]_{i}\mbf y}{\left[\diag{\widehat{\mbf s}_{\mathcal{B}_j}}\mathbf{D}_{\mathcal{B}_j}\left(\mathbf{I}-\mathbf{D}_{-\mathcal{A}_j}^T\left(\mbf D_{-\mca{A}_j}\mbf D_{-\mca{A}_j}^T\right)^{-1}\mbf D_{-\mca{A}_j}\right)\right]_{i}\mbf D_{\mca{A}_j}^T \widehat{\mbf s}_{\mca{A}_j}},
    \end{align}}
    which forms $|\mbf L_{j+1}|-1$ rows of $\mbf A_{j+1}$.
    
    Lastly, the decision of adding or removing a coordinate in part (c) is required. The condition for the joining time is $\lambda_{j+1}^{\mathrm{join}}\geq\lambda_{j+1}^{\mathrm{leave}}$, which can be expressed as
    {\footnotesize
    \begin{align}\label{gam.hit.leave.proj2}
        &\hspace{-1cm} \frac{\left[\left(\mbf D_{-\mca{A}_j}\mbf D_{-\mca{A}_j}^T\right)^{-1}\mbf D_{-\mca{A}_j}\right]_{\widehat{\tau}_{j+1}^{\,\textrm{join}}}\mbf y}{\widehat{s}_{j+1}^{\,\textrm{join}}+\left[\left(\mbf D_{-\mca{A}_j}\mbf D_{-\mca{A}_j}^T\right)^{-1}\mbf D_{-\mca{A}_j}\right]_{\widehat{\tau}_{j+1}^{\,\textrm{join}}}\mbf D_{\mca{A}_j}^T \widehat{\mbf s}_{\mca{A}_j}} \geq 
        \\[10pt]
        &\hspace{1.5cm} \frac{\left[\diag{\widehat{\mbf s}_{\mathcal{B}_j}}\mathbf{D}_{\mathcal{B}_j}\left(\mathbf{I}-\mathbf{D}_{-\mathcal{A}_j}^T\left(\mbf D_{-\mca{A}_j}\mbf D_{-\mca{A}_j}^T\right)^{-1}\mbf D_{-\mca{A}_j}\right)\right]_{\widehat{\tau}_{j+1}^{\,\textrm{leave}}}\mbf y}{\left[\diag{\widehat{\mbf s}_{\mathcal{B}_j}}\mathbf{D}_{\mathcal{B}_j}\left(\mathbf{I}-\mathbf{D}_{-\mathcal{A}_j}^T\left(\mbf D_{-\mca{A}_j}\mbf D_{-\mca{A}_j}^T\right)^{-1}\mbf D_{-\mca{A}_j}\right)\right]_{\widehat{\tau}_{j+1}^{\,\textrm{leave}}}\mbf D_{\mca{A}_j}^T \widehat{\mbf s}_{\mca{A}_j}}. 
    \end{align}}
    In the case of leaving time, which corresponds to $\lambda_{j+1}^{\mathrm{join}}<\lambda_{j+1}^{\mathrm{leave}}$, the sign of inequality in \eqref{gam.hit.leave.proj2} will flip. Observe that the decision on whether to add or remove a coordinate will only add one row to $\mbf A_{j}$. Also, it is important to note that $\mbf q= \mbf 0$ in all the above steps. 
\end{enumerate}

\subsection{Polyhedron Form After the Stopping Rule}
\label{sec:polyhedron.stopping.rule.proj2}
In the preceding section, we described how to construct the polyhedron matrix $\mbf A$ and vector $\mbf q$ after running the PRUTF algorithm for a fixed number of steps. However, the PRUTF algorithm, as discussed in Section \ref{sec:stopping.proj2}, terminates using a stopping rule. This stopping rule is developed based on the stochastic term of the dual variables, $\widehat{u}_{\mca A}^{\,\trm{st}}(t)= \left[\left(\mathbf{D}_{-\mathcal{A}}\mathbf{D}_{-\mathcal{A}}^T\right)^{-1}\mathbf{D}_{-\mathcal{A}}\right]_t\mathbf{y}$, for $t\in \{1, \ldots, m\} \backslash \mca A$. In the following, we will show that this stopping criterion can also be expressed as a polyhedron in $\mbf y$.

The stopping rule states that  Algorithm \ref{alg:tf.dual.path} stops upon the time the inequality
\begin{align}\label{stop.rule.proj2}
    \max_{0\leq t\leq 1} ~ \Big| \, \widehat{\mbf u}_{_{-\mca A}}^{\,\trm{st}} \big(\lfloor\, kt\,\rfloor \big) \, \Big| ~\leq~ \sigma\, x_{_\alpha} \big( k-r \big)^{(2r+1)/2}\,, \qquad \text{ for} \quad \, k=m-|\mca A|\,,
\end{align}
is satisfied. In order to show that this stopping rule creates a polyhedron, note that the condition \eqref{stop.rule.proj2} can be rewritten as
\begin{align}
    &\left[\left(\mathbf{D}_{-\mathcal{A}}\mathbf{D}_{-\mathcal{A}}^T\right)^{-1}\mathbf{D}_{-\mathcal{A}}\right]_t\mathbf{y} ~\geq~ -\sigma\, x_\alpha \big(k-r \big)^{(2r+1)/2}, \qquad\qquad\textrm{for} \quad t\notin \mca A 
    \\[11pt]
    -&\left[\left(\mathbf{D}_{-\mathcal{A}}\mathbf{D}_{-\mathcal{A}}^T\right)^{-1}\mathbf{D}_{-\mathcal{A}}\right]_t\mathbf{y} ~\geq~ -\sigma\, x_\alpha \big( k-r \big)^{(2r+1)/2}, \qquad\qquad \textrm{for} \quad t\notin \mca A.
\end{align}
These conditions append $2k$ rows to the matrix $\mbf A$ at each step. Additionally, the above conditions add $2k$ non-zero values $-\sigma\, x_\alpha \big( k-r \big)^{(2r+1)/2}$ to $\mbf q$.

\begin{remark}
We have just shown that the event $\big\{ \widehat{\bsy\tau}=\bsy\tau\, ,\, \widehat{\mbf s}=\mbf s \big\}$, for all fixed $\bsy\tau$ and $\mbf s$, constitutes a polyhedron of the form $\big\{\, \mbf y:~ \mbf A\, \mbf y \geq \mbf q \big\}$. It would also be interesting to only characterize change point set $\big\{ \widehat{\bsy\tau}= \bsy\tau \big\}$ in the form of a polyhedron. It turns out that $\big\{ \widehat{\bsy\tau}= \bsy\tau \big\}$ can be represented as the union of such polyhedra. More precisely, 
\begin{align}
    \Big\{ \mbf y:~ \widehat{\bsy\tau} (\mbf y)= \bsy\tau \Big\}= \bigcup_{\mbf S}\, \Big\{ \mbf y:~ \widehat{\bsy\tau} (\mbf y) =\bsy\tau\, ,\, \widehat{\mbf s} (\mbf y) =\mbf s \Big\}=\, \bigcup_{\mbf S}\, \Big\{ \mbf y:~ \mbf A_{{\,\mbf s}}\, \mbf y \geq \mbf q_{{\,\mbf s}} \Big\},
\end{align}
where the union is over all sign vectors, $\mbf s \in \{-1,\, 1\}^{^{\,|\row{\mbf A}|}}$. Observe that the number of elements for this union is $2^{^{\,|\row{\mbf A}|}}$ which can grow fast and become intractable when $|\row{\mbf A}|$ is moderately large.

\end{remark}

\section{Post-Detection Inference}
\label{sec:post.detection.proj2}

Having detected change points using the PRUTF algorithm, an appealing follow-up goal would be performing statistical inference on the significance of the changes at these locations. In this section, we provide inference tools to apply after implementing the PRUTF algorithm using the post-selection inference framework. Given the vector of observations $\mbf y$, we assume that $\widehat{\bsy\tau}=\big\{ \widehat{\tau}_{_1}, \,\ldots,\, \widehat{\tau}_{_J} \big\}$ is the set of ordered change points, $1 < \widehat{\tau}_{_1} < \widehat{\tau}_{_2} <\, \ldots\, < \widehat{\tau}_{_J} < n-r-1$, detected using the PRUTF algorithm. Additionally, assume that $\widehat{\mbf s}= \big\{\widehat{s}_1,\, \ldots,\, \widehat{s}_J \big\}$ is the set of signs associated with $\widehat{\bsy\tau}$. For notational convenience, we denote $\widehat{\tau}_0=0$ and $\widehat{\tau}_{_{J+1}}=n-r$. Our focus is on conducting valid statistical inference for the significance of changes at the locations $\widehat{\tau}_j$, $j=1\, ,\, \ldots\, ,\, J$.

Hypothesis tests about the significance of the detected change points can be cast in a general linear form $H_0:\, \bsy\eta^T\mbf f=0$, for some nonzero contrast vector $\bsy\eta\in \mbb R^{n}$. Of particular interest in this paper is the hypothesis $H_0:\, \mbf D_{\widehat{\tau}_j} \mbf f=0$, which tests the significance of change exactly at estimated location $\widehat{\tau}_j$, where $\mbf D_{\widehat{\tau}_{_j}}$ is the $\widehat{\tau}_{_j}$-th row of the penalty matrix $\mbf D$. However, our inferential framework is not specific to the choice $\bsy\eta^T= \mbf D_{\widehat{\tau}_j}$. 
Many other types of contrast vectors are possible, as long as $\bsy\eta$ is fixed by conditioning on the detection procedure. For example, the segment contrast, proposed in \cite{hyun2018exact}, considers the difference between averages of two neighboring segments $\big( \widehat{\tau}_{j-1}\, ,\, \widehat{\tau}_j \big]$ and  $\big( \widehat{\tau}_j\, ,\, \widehat{\tau}_{j+1} \big]$. Also, \cite{jewell2019testing} have used a window contrast of size $h$, which considers the difference between averages of $h$ consecutive points just before the estimated change point, i.e., $\big( \widehat{\tau}_{j}\, -\, h\, ,\, \widehat{\tau}_j \big]$ and $h$ consecutive points just after that, i.e.,  $\big( \widehat{\tau}_j\, ,\, \widehat{\tau}_{j}\, +\, h \big]$. The window contrast is suitable for checking whether a true change point exists near $\widehat{\tau}_j$. It is worth mentioning that the choice of $\bsy\eta$ ultimately depends on the objective of the practitioner.

Data dependence nature of change point detection methodologies leads to a random change point set $\widehat{\bsy\tau}= \widehat{\bsy\tau}\, (\mbf y)$. Associated with this randomly chosen $\widehat{\bsy\tau}$ is the vector $\bsy\eta= \bsy\eta(\widehat{\bsy\tau})$ which, in turn, is a random object. This randomness invalidates classical theory for conducting statistical inference about $\bsy\eta^T \mbf f$; see \cite{berk2013valid} for a thorough discussion. In such cases, post-detection inference allows us to carry out our analysis. In particular, post-detection inference revolves around the conditional distribution of $\bsy\eta^T \mbf y$ conditional on the selected change points. This conditioning makes $\bsy\eta= \bsy\eta(\widehat{\bsy\tau})$ to become a fixed vector. Now, the goal is to test a hypothesis that controls the conditional type-I error rate at level $\alpha$ as well as to build a conditional confidence interval $I_{\bsy\eta}$, such that
\begin{align*}
    \Pr \Big( \bsy\eta^T \mbf f \in I_{\bsy\eta}\, \Big|\, \widehat{\bsy\tau}= \bsy\tau \Big)\, \geq\, 1-\alpha,
\end{align*}
for all fixed $\bsy\tau$.

In a change point detection setting, \cite{hyun2018exact, hyun2018post} have exploited post-selection inference to compute p-values for the significance of change points found by fused lasso, Binary Segmentation, Wild Binary Segmentation, Circular Binary Segmentation. To boost the power of tests, \cite{jewell2019testing} have suggested a post-detection approach which attempts to reduce the size of conditioning events. This approach covers change points detected in a piecewise constant model with identical and independent Gaussian random noises using a $\ell_0$-penalization. In \cite{duy2020computing}, the authors have also considered post-selection inference for change points estimated by using a dynamic programming for a $\ell_0$-penalization. 

Conducting inference for $\bsy\eta^T \mbf f$ in the post-detection framework requires the knowledge about the conditional distribution of $\bsy\eta^T \mbf y$  given $\{\widehat{\bsy\tau}= \bsy\tau\}$. As shown in Section \ref{sec:polyhedron.proj2}, $\{\widehat{\bsy\tau}= \bsy\tau\, ,\, \widehat{\mbf s}= \mbf s \}$ creates the polyhedron of the form $\big\{\, \mbf y:~ \mbf A\, \mbf y \geq \mbf q \big\}$, and $\{\widehat{\bsy\tau}= \bsy\tau \}$ is a union of such polyhedra over all possible sign vectors $\mbf s$. Therefore, it is easier to obtain the conditional distribution of $\bsy\eta^T \mbf y$ given $\{\widehat{\bsy\tau}= \bsy\tau\, ,\, \widehat{\mbf s}= \mbf s \}$, a single polyhedron. Observe that inferences that are valid conditional on this finer event will also be valid conditional on $\{\widehat{\bsy\tau}= \bsy\tau\}$, \citep{lee2016exact}. In what follows, we illustrate techniques to compute valid post-detection $p$-value for the hypothesis $H_0:\, \bsy\eta^T\mbf f=0$ and to construct a post-detection confidence interval for the parameter $\bsy\eta^T\mbf f$.

Ignoring the post-detection framework for a moment, recall that when the change points are assumed fixed, the inference about $\bsy\eta^T \mbf f$, depending on whether the error variance is known or unknown, is based on the normal or $t$ distributions, respectively. More specifically, when $\sigma^2$ is known, the statistic $Z=\bsy\eta^T \mbf y/\sigma \|\bsy\eta\|$ and when $\sigma^2$ is unknown, the statistic $T=\bsy\eta^T \mbf y/\widehat{\sigma} \|\bsy\eta\|$, are employed to make inference for $\bsy\eta^T \mbf f$. We will essentially use the same statistics $Z$ and $T$ in post-detection inference and focus on determining their respective conditional distributions. We emphasize that these distributions are no longer the usual normal or $t$ distributions as they must be conditioned on the detected change points.

To define our proposed test statistics, again assume that the PRUTF algorithm has detected $J$ change points at locations $\widehat{\bsy\tau}= \big\{ \widehat{\tau}_{_1},\,\ldots,\,\widehat{\tau}_{_J} \big\}$ with the corresponding signs $\widehat{\mbf s}= \big\{\widehat{s}_{_1},\, \ldots,\, \widehat{s}_{_J} \big\}$. These change points partition the entire signal $\mbf f$ into $J+1$ segments, with each segment having its distinct polynomial signal of order $r$, namely $\mbf f_{j}$, $j=0\, ,\, \ldots\, ,\, J$. Also, let $\mbf y_{_j}$ denote the sub-vector of observations corresponding to the $j$-th segment; thus,
\begin{align*}
    \mbf f^T\mbf y=\ssum[j]{0}{J}\mbf f_{j}^T\mbf y_{_j}.
\end{align*}
Implementing the least square approach to estimate $\mbf f_{j}$ results in $\widehat{\mbf f}_{j}= \left(\mbf X_{j}^T\mbf X_{j}\right)^{-1}\mbf X_{j}^T\mbf y_{_j}$, for $j=0,\, \ldots,\, J$,
where $\mbf X_j$ is defined as
\begin{align}\label{designmat.segment.proj2}
    \mbf X_{j}=\begin{pmatrix}
      1 & \frac{\widehat{\tau}_{_j}+1}{n} & \left(\frac{\widehat{\tau}_{_j}+1}{n}\right)^2 & \cdots & \left(\frac{\widehat{\tau}_{_j}+1}{n}\right)^r 
      \\[10pt]
      1 & \frac{\widehat{\tau}_{_j}+2}{n} & \left(\frac{\widehat{\tau}_{_j}+2}{n}\right)^2 & \cdots & \left(\frac{\widehat{\tau}_{_j}+2}{n}\right)^r \\
      \vdots & \vdots & \vdots &  & \vdots \\
      1 & \frac{\widehat{\tau}_{_{j+1}}}{n} & \left(\frac{\widehat{\tau}_{_{j+1}}}{n}\right)^2 & \cdots & \left(\frac{\widehat{\tau}_{_{j+1}}}{n}\right)^r \\
    \end{pmatrix}.
\end{align}
In words, $\mbf X_j$ is the design matrix of the $r$-th polynomial regression of $\mbf y_{_j}$ on the indices of $j$-th segment, $\widehat{\tau}_{_j}+1,\,\ldots,\,\widehat{\tau}_{_{j+1}}$.
We also denote the projection matrix onto the column space of $\mbf X_j$ as $\mbf P_j=\mbf X_j\left(\mbf X_j^T\mbf X_j\right)^{-1}\mbf X_j^T$. Observe that $\mbf P_j\, \mbf f_j= \mbf f_j$, for $j=0\,,\,\ldots\, ,\, J$, or equivalently $\mbf P\, \mbf f= \mbf f$, where $\mbf P$ is a block diagonal matrix whose diagonal entries are the submatrices. With these notations, an unbiased estimator of the error variance, when $\sigma^2$ is unknown, is given by 
\begin{align}
    \widehat{\sigma}^2=\frac{1}{d}\,\ssum[j]{0}{J}\, \big\|(\mbf I_j-\mbf P_j)\,\mbf y_{_j} \big\|^2=\, \frac{1}{d}\, \big\|(\mbf I-\mbf P)\,\mbf y \big\|^2\,,
\end{align}
where $d=n-(J+1)$. 

According to the Gaussian model of \eqref{fmodel.proj2}, $\mbf y$ follows the exponential family of the form
\begin{align*}
    \mbf y 
    &\sim\exp\left\{ \frac{1}{\sigma^2}\, \mbf f^T\mbf y-\frac{1}{2\sigma^2}\|\mbf y\|^2-\frac{\|\mbf f\|^2}{2\sigma^2} \right\}.
\end{align*}
Decomposing the data into the direction of $\bsy\eta$ and orthogonal to $\bsy\eta$, as well as using the fact that $\mbf P\, \mbf f=\mbf f$, yield
\begin{align}\label{expfam.form.proj2}
    \mbf y &\sim\exp\left\{ \frac{1}{\sigma^2}\, \mbf f^T \big(\mbf P_{\eta}+\mbf P-\mbf P_{\eta} \big)\, \mbf y- \frac{1}{2\, \sigma^2}\, \|\mbf y\|^2-\frac{\|\mbf f\|^2}{2\sigma^2} \right\}
    \\[8pt]
    &=\exp\left\{\left(\frac{1}{\sigma \|\bsy\eta\|}\right)\left(\bsy\eta^T\mbf f\right)^T \left(\frac{\bsy\eta^T \mbf y}{\sigma \|\bsy\eta\|}\right)+ \frac{1}{\sigma^2}\, \mbf f^T \big(\mbf P- \mbf P_{\eta})\, \mbf y-\frac{1}{2\, \sigma^2}\, \|\mbf y\|^2- \frac{\|\mbf f\|^2}{2\, \sigma^2} \right\},
\end{align}
where $\mbf P_{\eta}$ is the orthogonal projection on the space spanned by $\bsy\eta \in \mbb R^n$, defined as
\begin{align*}
   \mbf P_\eta=\bsy\eta\left(\bsy\eta^T\bsy\eta\right)^{-1} \bsy\eta^T=\frac{\bsy\eta \,\bsy\eta^T}{\|\bsy\eta \|^2}\,\cdot
\end{align*}
Consider a multi-parameter exponential family in which the vector of natural parameters can be split into two sub-vectors: the target and the nuisance parameters. The conditional distribution, given sufficient statistics associated with nuisance parameters, depends only on the target parameters. It is known that this conditional distribution belongs to an exponential family with the same target parameters, the same sufficient statistics, but a different reference measure and the normalizing constant. This fact allows us to derive the conditional distribution of the statistics $Z$ and $T$, given the detected change points, their signs and the sufficient statistics of the nuisance parameters. Consequently, we can perform statistical inference for the target parameter $\bsy\eta^T\mbf f$ based on the derived distributions. We divide our presentation into two parts: known error variance and unknown error variance to perform such inferences.

\subsection{Known Error Variance}
\label{sec:knownvar.proj2}
When the error variance $\sigma^2$ is known, the statement \eqref{expfam.form.proj2} reveals that $\bsy\eta^T \mbf y$ and  $\mbf  V= \big( \mbf P-\mbf P_{\eta} \big) \,\mbf y$ are sufficient statistics for $\bsy\eta^T \mbf f$ and the nuisance parameters, respectively. As previously explained, the conditional distribution of $\bsy\eta^T \mbf y$ given $\mbf  V$ eliminates nuisance parameters from our analysis. Therefore, when $\sigma^2$ is known, we base our analysis on the statistic
\begin{align}\label{zstat.proj2}
    Z=\frac{\bsy\eta^T \mbf y}{\sigma\,  \|\bsy\eta\|}\,\cdot
\end{align}
We further seek to specify the conditional distribution of this statistic given $\left\{\, \mbf  V,\, \mbf A\mbf y\geq \mbf q \right\}$. Recall that the polyhedron $\left\{\mbf A \mbf y \geq \mbf q \right\}$ is the substitute for $\big\{ \widehat{\bsy\tau}\, ,\, \widehat{\mbf s} \big\}$, estimated using PRUTF. The following theorem illustrates that this conditional distribution is indeed a truncated normal distribution with an explicitly specified truncation set.

\begin{theorem}\label{thm:ztrunc.proj2}
Suppose $\mbf y$ follows model \eqref{fmodel.proj2}, where $\sigma^2$ is assumed known. For a nonzero contrast vector $\bsy\eta$,
\begin{enumerate}[label=\alph*)]
    \item the conditional distribution of $Z$ in \eqref{zstat.proj2}, given
    \begin{align}\label{ztrunc.law.proj2}
        \Big\{\, \mbf  V,\, \mbf A\mbf y\geq \mbf q \Big\}, 
    \end{align}
    is a normal distribution truncated to the interval $\big[\mca V_Z^-\,,\,\mca V_Z^+ \big]$, provided $\mca V_Z^0\geq 0$. This distribution is denoted by ${\rm TN}\big( \bsy\eta^T \mbf f,\, 1,\, [\mca V_Z^-,\mca V_Z^+] \big)$. The truncation boundaries $\mca V_Z^-=\mca V_Z^- \big(\mbf V \big)$, $\mca V_Z^+=\mca V_Z^+ \big(\mbf V \big)$ and $\mca V_Z^0=\mca V_Z^0 \big(\mbf V \big)$ are given by
    {\small
    \begin{align}\label{ztrunc.bounds.proj2}
       \mca V_Z^-&=\max_{i:\,\rho_{_i}>0}\,\frac{\big[ \mbf q - \mbf A \mbf  V\big]_i}{\sigma\rho_{_i}}\,,
       \qquad
       \mca V_Z^+= \min_{i:\,\rho_{_i}<0}\, \frac{\big[ \mbf q - \mbf A \mbf  V\big]_i}{\sigma\rho_{_i}}\,,
       \qquad
       \mca V_Z^0= \min_{i:\,\rho_{_i}=0}\,  \big[ \mbf A \mbf  V - \mbf q \big]_i \,,
    \end{align}}
     where $\rho_{_i}=\big[ \mbf A\bsy\eta/\|\bsy\eta\| \big]_i$, for $i=0\, ,\, 1\, ,\, \ldots\, ,\, \big| \row{\mbf A} \big|$. Observed that $\big| \row{\mbf A} \big|$ denotes the number of rows in matrix $\mbf A$.
     
     \item Moreover, let $\Phi^{\,[a\, ,\, b]} \, (\cdot)$ be the cumulative distribution function of $\,{\rm TN} \big( 0,\, 1,\, [a\, ,\, b] \big)$, thus, under the null hypothesis $H_0:\,\bsy\eta^T \mbf f=0$, the conditional distribution of $1-\Phi^{[\mca V_Z^-,\mca V_Z^+]}\,(Z)$ given the event $\big\{\, \mbf A\mbf y\geq \mbf q \big\}$ is uniform on the unit interval $[0,\,1]$, that is, 
     \begin{align}\label{tn.stat.proj2}
        1-\Phi^{[\mca V_Z^-,\mca V_Z^+]}\,(Z)\, \Big |\, \big\{\, \mbf A\mbf y\geq \mbf q \big\}\sim {\rm U}(0\, ,\, 1)\,.
     \end{align}
     We refer to this statistic as ${\rm TN}$ statistic.
\end{enumerate}
\end{theorem}
The proof is given in Appendix \ref{prf:ztrunc.proj2}.

Theorem \ref{thm:ztrunc.proj2} enables us to compute a post-detection {\it p}-value for the hypothesis $H_0:\,\bsy\eta^T \mbf f=0$ as well as to construct a post-detection confidence interval for $\bsy\eta^T \mbf f$. These tasks can be carried out using the ${\rm TN}$ statistic in \eqref{tn.stat.proj2}. In particular, for the two-sided hypothesis testing problem
\begin{align}\label{twoside.hypo.proj2}
    H_0: \bsy\eta^T\mbf f=0\qquad\qquad\textrm{vs}\qquad\qquad H_1: \bsy\eta^T\mbf f \neq 0,
\end{align}
the value of $2 \min\left\{1-\Phi^{[\mca V_Z^-,\mca V_Z^+]}\,(Z)\, ,\, \Phi^{[\mca V_Z^-,\mca V_Z^+]}\,(Z)\right\}$ serves as a valid post-detection $p$-value, since under the null hypothesis
\begin{align*}
    \Pr \bigg( 2 \min\left\{1-\Phi^{[\mca V_Z^-,\mca V_Z^+]}\,(Z) \, ,\, \Phi^{[\mca V_Z^-,\mca V_Z^+]}\,(Z)\right\}\leq \alpha\, \bigg| \, \mbf A\mbf y\geq \mbf q \bigg)= \alpha,
\end{align*}
for all $0\leq\alpha\leq 1$. To construct a two-sided post-detection confidence interval, define the confidence limits $L_{_Z}(Z)$ and $U_{_Z}(Z)$ such that \begin{align}\label{ztrunc.ci.proj2}
    1-\Phi^{[\mca V_Z^-,\mca V_Z^+]}\,\left(\frac{\bsy\eta^T \mbf y-L_{_Z}(Z)}{\sigma\|\bsy\eta\|}\right)=\frac{\alpha}{2},
    \qquad \text{ and }\qquad
    \Phi^{[\mca V_Z^-,\mca V_Z^+]}\,\left(\frac{\bsy\eta^T \mbf y-U_{_Z}(Z)}{\sigma\|\bsy\eta\|}\right)=\frac{\alpha}{2}\,. 
\end{align}
These limits are well characterized since the survival function of ${\rm TN} \big( \mu,\, 1,\, [a ,\, b] \big)$ monotonically increases with respect to $\mu$. Hence, the interval $\big[ L_{_Z}(Z)\,,\,U_{_Z}(Z) \big]$ is a valid two-sided post-detection confidence interval at a $1-\alpha$ level for $\bsy\eta^T \mbf f$. This confidence interval can be interpreted in the following manner. If $\mbf y$ is repeatedly drawn from model \eqref{fmodel.proj2} and the PRUTF algorithm is run, among those cases in which $\big\{ \widehat{\bsy\tau}\, ,\, \widehat{\mbf s} \big\}$ are detected, the interval $\big[ L_{_Z}(Z)\,,\,U_{_Z}(Z) \big]$ contains the parameter $\bsy \eta^T \mbf f$ with a relative frequency approaching $1-\alpha$.

\begin{remark}\label{ztrunc.onesided.ci.proj2}
The PRUTF algorithm estimates sings of change points in addition to their locations. We can incorporate this knowledge in forming the alternative hypothesis, that is, $H_1:\, \widehat{s}_{_{\widehat{\tau}_{_j}}} \mbf D _{_{\widehat{\tau}_{_j}}} \mbf f >0$. This alternative hypothesis means that $\widehat{\tau}_{_j}$ is a significant change point whose jump is in the direction of $\, \widehat{s}_{_{\widehat{\tau}_{_j}}}$. Note that this test is more powerful than its two-sided counterpart. \cite{tibshirani2016exact} have provided a comparison between one-sided and two-sided tests for the significance of the selected variables using lasso.
In general, suppose we are interested in testing the one-sided hypothesis $H_0:\, \bsy\eta^T\mbf f=0$ against $H_1:\, \bsy\eta^T\mbf f > 0$. As with the two-sided hypotheses, $1-\Phi^{[\mca V_Z^-,\mca V_Z^+]}\,(Z)$ is a valid post-detection $p$-value for the one-sided test. Additionally, $\big[ L_{_Z}(Z)\, ,\, \infty \big)$ is a one-sided post-detection confidence interval for $\bsy\eta^T\mbf f$ where 
\begin{align*}
    1-\Phi^{\,[\mca V_Z^-,\mca V_Z^+]}\,\left(\frac{\bsy\eta^T \mbf y-L_{_Z}(Z)}{\sigma\|\bsy\eta\|}\right)=\alpha.
\end{align*}
\end{remark}


\subsection{Unknown Error Variance}
\label{sec:unknownvar.proj2}
This section concerns post-detection inference after estimating change points, in the more realistic case, when the error variance $\sigma^2$ is unknown. The main methods proposed for post-selection inference such as in \cite{lee2016exact}, \cite{tibshirani2016exact}, and \cite{ taylor2018post} proceed with a known $\sigma^2$, with the exception of \cite{fithian2015selective}. In the case of an unknown $\sigma^2$, we must further condition our inference on the sufficient statistic associated with the nuisance parameter $\sigma^2$.  

According to \eqref{expfam.form.proj2}, in the case of an unknown $\sigma^2$, the term $\bsy\eta^T\mbf y$ is the sufficient statistic for the target parameter $\bsy\eta^T\mbf f/\sigma^2$ and $\big( \mbf V\, ,\, \|\mbf y\|^2 \big)$ is a joint sufficient statistic for the nuisance parameters. Since testing $\bsy\eta^T\mbf f/\sigma^2=0$ is equivalent to testing $\bsy\eta^T\mbf f=0$, hence, we construct our analysis based on the statistic 
\begin{align}\label{tstat.proj2}
    T=\frac{\bsy\eta^T \mbf y} {\widehat{\sigma}\, \|\bsy\eta\|},
\end{align}
where $\widehat{\sigma}^2=d^{-1}\, \big\|(\mbf I-\mbf P)\,\mbf y \big\|^2$, with $d=n-(J+1)$. Notice that $\widehat{\sigma}^2$ is simply a pooled estimate of the error variance $\sigma^2$ using $J+1$ segments created by the detected change points $\widehat{\bsy\tau}= \big\{ \widehat{\tau}_{_1},\,\ldots,\,\widehat{\tau}_{_J} \big\}$. The next step is to find the conditional distribution of $T$, given the sufficient statistics associated with the nuisance parameters and the polyhedron event identified by the detection procedure. Clearly, this statistic is distributed as a $t$ distribution constrained to the set $\big\{\, \mbf  V,\, \|\mbf y\|^2,\, \mbf A\mbf y\geq \mbf q \big\}$. We establish the corresponding distribution in the following theorem, whose proof is given in Appendix \ref{prf:ttrunc.proj2}.

\begin{theorem}\label{thm:ttrunc.proj2}
Suppose $\mbf y$ follows model \eqref{fmodel.proj2} and $\sigma^2$ is assumed unknown. For a nonzero contrast vector $\bsy\eta$,
\begin{enumerate}[label=\alph*)]
    \item the conditional distribution of $T$ given
    \begin{align}\label{ttrunc.law.proj2}
    \Big\{\, \mbf  V,\, \big\| \mbf y \big\|^2\, ,\, \mbf A \mbf y \geq \mbf q \Big\}, 
    \end{align}
     is a location-scale $t$ distribution with mean $\bsy\eta^T \mbf f$, variance 1 and degrees of freedom $d=n-(J+1)$, truncated to the interval $[\mca V_T^-\,,\,\mca V_T^+]$, denoted by ${\rm Tt} \big( \bsy\eta^T \mbf f ,\, 1,\, d,\, [\mca V_T^-,\mca V_T^+] \big)$. The truncation boundaries $\mca V_T^-=\mca V_T^- \big(\mbf V,\, W \big)$ and $\mca V_T^+=\mca V_T^+ \big( \mbf V,\, W \big)$ are given by
    {\small
    \begin{align}\label{ttrunc.bounds.proj2}
    \Big[ \mca V_T^-\, , \, \mca V_T^+ \Big] = \bigcap_{i=1}^{ |\row{\mbf A}|} \Bigg\{ t\in \mbb R :
    \Big[ \mbf A \mbf V - \mbf q \Big]_i\, t^2 &+ \bigg( 2\, \frac{\bsy\eta^T(\mbf I -\mbf P)\, \mbf y}{\widehat{\sigma} \|\bsy\eta\|}\, \Big[ \mbf A \mbf V - \mbf q \Big]_i + \frac{W \rho_i}{\widehat{\sigma}} \bigg)\,t
    \\[8pt]
    &+\, \Big[ \mbf A \mbf V - \mbf q \Big]_i\, d  \geq 0 \Bigg\},
    \end{align}}
     where $W= \big\| \big(\mbf I-\mbf P+\mbf P_{\eta} \big)\, \mbf y \big\|^2$.
     
    \item In addition, let $G_d^{\,[a,\, b]} (\cdot)$ denote the cumulative distribution function of ${\rm Tt} \big( 0\, ,\, 1\, ,\, d\,$ ,$\, [a\, ,\, b] \big)$, then under the null hypothesis
     \begin{align}\label{tt.stat.proj2}
         1-G_d^{\,[\mca V_T^{-} ,\, \mca V_T^{+}]}\,(T)\, \Big|\, \big\{ \mbf A\mbf y\geq \mbf q \big\} \sim U(0\, ,\, 1).
     \end{align}
     We refer this statistic as the {\rm Tt} statistic.
\end{enumerate}
\end{theorem}
Theorem \ref{thm:ttrunc.proj2} allows us to perform post-detection tests and construct  post-detection confidence intervals for $\bsy\eta^T \mbf f$. In the same fashion as the truncated normal case, the quantity
\begin{align*}
    2\min\left\{1-G_d^{\,[\mca V_T^{-} ,\, \mca V_T^{+}]} ~ (T) ~ ,\, G_d^{\,[\mca V_T^{-} ,\, \mca V_T^{+}]}\,(T)\right\}
\end{align*}
is a post-detection $p$-value for the two-sided hypothesis problem \eqref{twoside.hypo.proj2}. A post-detection confidence interval for $\bsy\eta^T\mbf f$, when $\sigma^2$ is unknown, is also given by $\big[ L_T(T)\,,\, U_T(T) \big]$, where
\begin{align}\label{ttrunc.ci.proj2}
    1-G_d^{[\mca V_Z^-,\mca V_Z^+]}\,\left(\frac{\bsy\eta^T \mbf y-L_T(T)}{\widehat{\sigma}\|\bsy\eta\|}\right)=\alpha/2 \,, 
    \qquad \trm{and} \qquad
    G_d^{[\mca V_Z^-,\mca V_Z^+]}\,\left(\frac{\bsy\eta^T \mbf y-U_T(T)}{\widehat{\sigma}\|\bsy\eta\|}\right)= \alpha/2 \,. 
\end{align}
The same technique explained in Remark \ref{ztrunc.onesided.ci.proj2} can be used to construct a one-sided confidence interval for $\bsy\eta^T\mbf f$.

\begin{figure}[!ht]
\begin{subfigure}{.5\textwidth}
  \centering
  \includegraphics[width=1\linewidth]{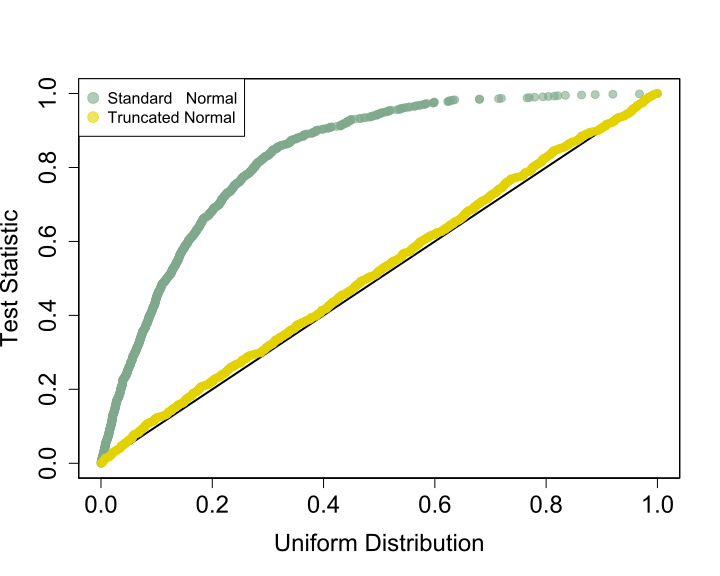}
  \caption{Q-Q plot of survival function of $Z$.}
\end{subfigure}
\begin{subfigure}{.5\textwidth}
  \centering
  \includegraphics[width=1\linewidth]{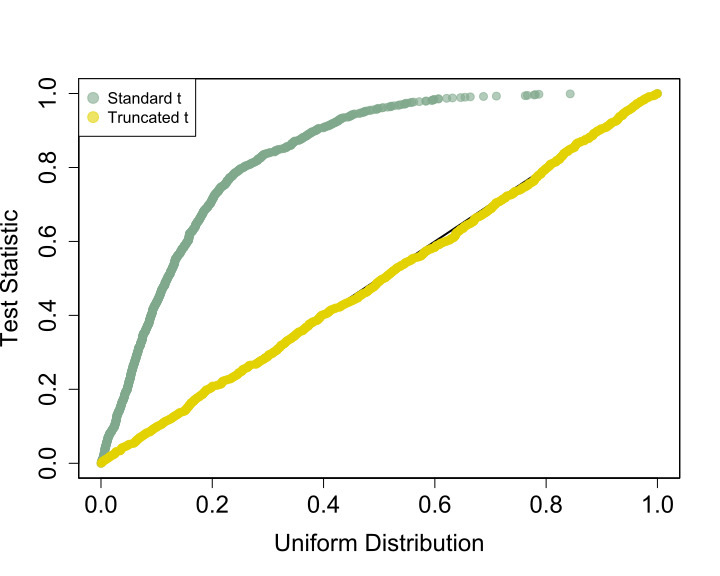}
  \caption{Q-Q plot of survival function of $T$.}
\end{subfigure}
\caption{ The Q-Q plots of survival functions of $Z$ and $T$ for standard and truncated normal and t distributions. A piecewise constant signal of size $n=100$ and a true change point at $\bsy\tau= \{50\}$ is considered. The left panel displays quantile-quantile plot, constructed over 1000 simulations, of the distribution of the survival function of $Z$ in \eqref{tn.stat.proj2} for the first detected change point, under standard normal (green dots) and truncated normal (yellow dots) distributions. The right panel shows Q-Q plots of the distribution of the survival function of $T$ in \eqref{tt.stat.proj2} for the first detected change point, under t distribution (green dots) and truncated t distribution (yellow dots).}
\label{fig:ZT_qqplot}
\end{figure}

Figure \ref{fig:ZT_qqplot} demonstrates the distribution of TN and Tt statistics, stated in \eqref{tn.stat.proj2} and \eqref{tt.stat.proj2}, by displaying their quantiles versus those of $U(0\, ,\, 1)$. The figure also represents the distribution of the two statistics when the truncated normal and truncated t distributions in \eqref{tn.stat.proj2} and \eqref{tt.stat.proj2} are replaced with their untruncated counterparts. The figure certifies that the distribution of $Z$ and $T$ change from normal and t distributions to truncated normal and truncated $t$ distributions, respectively, while accounting for the detection procedure.

\begin{remark}
We would like to emphasize that the confidence intervals constructed by using \eqref{ztrunc.ci.proj2} and \eqref{ttrunc.ci.proj2} employ the knowledge of the conditional distributions given the estimated change points and their signs, $\big\{ \widehat{\bsy\tau}= \bsy\tau\, ,\, \widehat{\mbf s}= \mbf s \}$, and therefore, have to be interpreted conditionally. These confidence intervals  can also be interpreted simultaneously after applying the Bonferroni correction to properly adjust for multiplicity. That is, we would compute confidence intervals at the level $1-\textstyle\dfrac{\alpha}{\widehat{J}}\,$, where $\widehat{J}$ is the number of change points estimated by mPRUTF.

One approach for constructing simultaneous confidence intervals in post-detection framework is to use the universal valid post-selection inference of \cite{berk2013valid}, by considering all possible detection procedures. More precisely, for any change point detection procedure $\widehat{\bsy\tau}: \mbf y \longrightarrow \mca T$, where $\mca T$ is the collection of all possible change point sets, the goal of the universal post-selection inference, referred to as simultaneous PoSI, is to construct confidence intervals for $\bsy\eta_{j,\, \widehat{\tau}}^T \, \mbf f$. In this notation, $\bsy\eta_{j,\, \widehat{\tau}}$ denotes a nonzero contrast vector associated with $\widehat{\tau}_j \in \widehat{\bsy\tau}$.  According to the simultaneous PoSI method, a valid confidence interval for $\bsy\eta_{j,\, \widehat{\tau}}^T \, \mbf f$ takes the form
\begin{align*}
    \widehat{I}_{j,\, \widehat{\tau}}\, (K_{\alpha}) = \Big(\, \bsy\eta_{j,\, \widehat{\tau}}^T \, \mbf y ~\pm~ K_{\alpha}\, \sigma\, \| \bsy\eta_{j,\, \widehat{\tau}} \|\, \Big),
\end{align*}
where the constant $K_{\alpha}$ is derived such that 
\begin{align*}
    \Pr \Big(\, \bsy\eta_{j,\, \widehat{\tau}}^T \, \mbf f\, \in\, \widehat{I}_{j,\, \widehat{\tau}} (K_{\alpha})\, ,\quad \forall\,\, \widehat{\tau}_j \in \widehat{\bsy\tau}\, \Big)\, \geq \, 1-\alpha,
    \qquad\qquad
    \forall\,\, \widehat{\bsy\tau} \in \mca T.
\end{align*}
Note that the universal valid post-selection inference provides simultaneity both over the change points and over all detection procedures. This method has the advantage that it does not depend on the change point detection procedure. However, unless the number of possible change point sets is fairly small, this method is computationally challenging as the collection of all possible change point sets becomes intractable. Additionally, the confidence intervals derived from this method are unnecessarily wide as it disregards the knowledge of how the change points are estimated.
To improve the universal valid post-detection inference, a method based on a simultaneous over selection (SoS) criterion was proposed by \cite{zrnic2020post}. The method aims to construct confidence intervals for selected variables in a fixed  stable model selection procedure in the regression context. We note that computing simultaneous confidence intervals is not among the goals of our inference and refer interested readers to \cite{berk2013valid}, \cite{benjamini2019confidence}, \cite{andrews2019inference} and \cite{zrnic2020post}.
\end{remark}

\begin{remark}
We have thus far explained how to carry out statistical inference for the detected change points after completing the PRUTF algorithm. In other words, we have computed post-selection $p$-values and confidence intervals conditional on the entire set of estimated change points $\widehat{\bsy\tau}$. An alternative way is to perform statistical inferences for the change points in a sequential manner. In particular, suppose $\widehat{\bsy\tau}_{j-1}$ and  $\, \widehat{\mbf s}_{j-1}$ are the vectors of detected change points and their corresponding signs at the iteration $j-1$ of the PRUTF algorithm. Additionally, suppose $\widehat{\tau}_j$ is the detected change point at step $j$. In the sequential scheme, the interest is in making inference about the significance of change at the location $\widehat{\tau}_j$, given $\big\{ \widehat{\bsy\tau}_j\, ,\, \widehat{\mbf s}_j \big\}$. Notice that the tools provided in Theorems \ref{thm:ztrunc.proj2} and \ref{thm:ttrunc.proj2} are still applicable in the sequential scheme, but the matrix $\mbf A$ and vector $\mbf q$ in the polyhedron representation must be adjusted accordingly.  See \cite{lockhart2014significance} and \cite{tibshirani2016exact} which have pursued statistical inference in a sequential approach.
\end{remark}

\section{A Critique of Post-Detection Inference Methods}
\label{sec:critique.post.detection.proj2}

In post-detection inference, we lose part of the information in the data because it has already been used to estimate change points. As a result, the post-detection distributions change from normal or $t$ distributions to truncated normal or truncated $t$ distributions (Theorem \ref{thm:ztrunc.proj2} and Theorem \ref{thm:ttrunc.proj2}). The performance of post-detection inference heavily depends on the amount of information in the data used for change point detection. Over-conditioning, which uses too much information for the detection procedure and leaves little information for inference, leads to a significant loss in the power of tests and unacceptably wide confidence intervals; see \cite{fithian2014optimal}, \cite{lee2016exact} and \cite{kivaranovic2020length}. 

The over-conditioning issue in post-selection inference has motivated researchers to suggest various approaches that have higher left-over information for inference. One solution to the problem is data splitting or data carving in which the data is divided into two parts, one part for model selection and the other part for inference \citep{fithian2014optimal}. Another approach, put forward by \cite{tian2018selective}, is the idea of randomization, which selects the model based on not the actual dataset but a
noise-perturbed version. Also, \cite{liu2018more} have suggested classifying the selected model into high-value and low-value submodels. The post-selection inference will then be conducted by only conditioning on the high-value submodel. Although these approaches improve the performance of post-selection inference, they also share some drawbacks; see \cite{fithian2014optimal}. 

The problem of wide length confidence intervals in post-selection inference has specifically been investigated by \cite{kivaranovic2020length} for models chosen by lasso. They have established that a confidence interval produced by the truncated Gaussian distribution with a finite truncation boundary (either the upper or lower bound) always has an infinite expected length. In the next theorem, we will show that the same property also holds for a truncated $t$ distribution. More precisely, let a truncation set $\mca S$ to be the union of finitely many open intervals, where the intervals might be unbounded. In other words, $\mca S$ can be represented in the form of $\mca S= \bigcup_{i=1}^k \big(a_i\, ,\, b_i \big)$, for a finite value $k$, and for some $a_1 < b_1 < a_2 < b_2 < \ldots < a_k < b_k$ in $\mbb R$. 

\begin{theorem}\label{thm:cilength.zttrunc.proj2}

\begin{enumerate}[label=\alph*)]
    \item[]
    \item Let $Z \sim {\rm TN} \big(\mu,\, \sigma^2,\, \mca S_Z \big)$, where the truncation set $\mca S_Z$ is of the form $\mca S_Z=\bigcup_{\,i=1}^{\,k} \big( a_i\, ,\, b_i \big)$. Also, let $\big[ L_Z(Z)\, ,\, U_Z(Z) \big]$ be a conditional confidence interval for $\mu$, given $\mca S_Z$. Therefore, if the truncation set $\mca S_Z$ is bounded, either from below $(-\infty< a_1)$ or from above $(b_k< \infty)$, then
    \begin{align}\label{ztrunc.infinite.expect.length}
       E \Big[ U_Z \big( Z \big)- L_Z \big( Z \big ) \Big]= \infty.
    \end{align}
    
    \item Let $T \sim {\rm Tt} \big( \mu,\, \sigma^2,\, d,\, \mca S_T \big)$, where the truncation set $\mca S_T$ is of the form $\mca S_T=\bigcup_{\,i=1}^{\,m} \big(c_i\, ,\, d_i \big)$. Also, let $\big[ L_{_T}(T)\, ,\, U_{_T}(T) \big]$ be a conditional confidence interval for $\mu$, given $\mca S_T$. Similar to part (a), if the truncation set $\mca S_T$ is bounded, either from below $(-\infty< c_1)$ or from above $(d_m< \infty)$, then
    \begin{align}\label{ttrunc.infinite.expect.length}
       E \Big[ U_T \big( T \big)- L_T \big( T \big ) \Big]= \infty.
    \end{align}
\end{enumerate}


\end{theorem}

The proof is provided in Appendix \ref{prf:cilength.zttrunc.proj2}.

\begin{figure}[!t]
\begin{center}
\begin{subfigure}{.45\textwidth}
  \centering
  \includegraphics[width=1\linewidth]{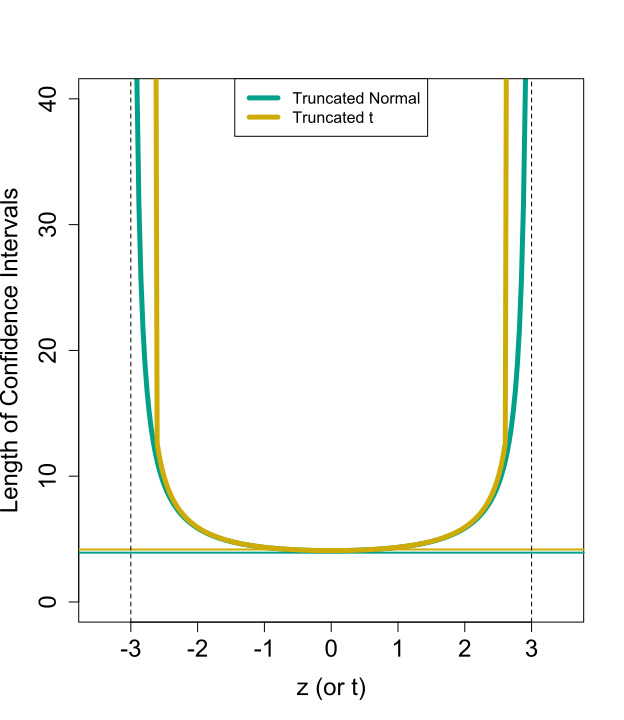}
  \caption{Truncation set $\mca S= [-3\, ,\, 3]$.}
  \label{fig:ci.length.bounded.proj2}
\end{subfigure}
\quad
\begin{subfigure}{.45\textwidth}
  \centering
  \includegraphics[width=1\linewidth]{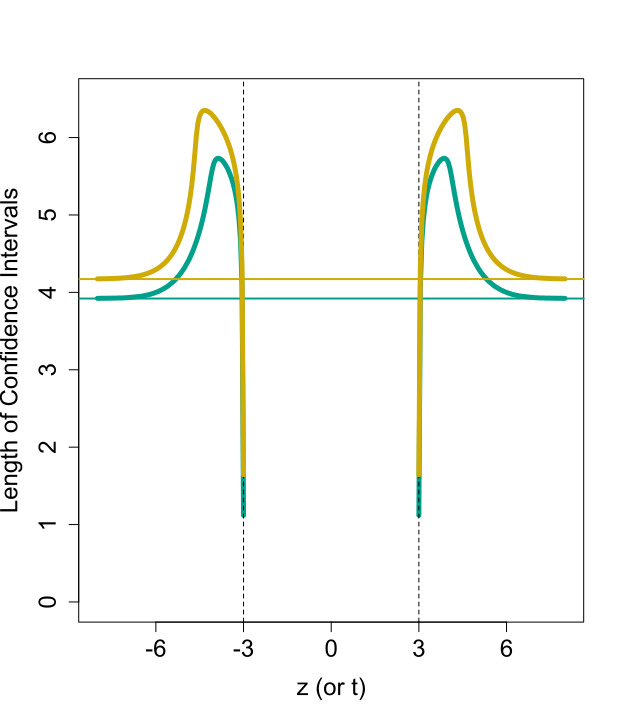}
  \caption{Truncation set $\mca S= (-\infty\, ,\, -3) \cup (3\, ,\, \infty)$.}
  \label{fig:ci.length.unbounded.proj2}
\end{subfigure}
\caption{ Lengths of confidence intervals for a truncated normal distribution and a truncated t distribution with a bounded truncation set (left panel) and an unbounded truncation set (right panel). The solid horizontal lines in both panels show the length of confidence intervals (at the level $0.95$) for the usual (untruncated) normal and t distributions.}
\label{fig:ci.length.proj2}
\end{center}
\end{figure}

Theorem \ref{thm:cilength.zttrunc.proj2} states that the truncation set is crucial in constructing a desirable confidence interval in post-selection inference. To figure out why post-selection confidence intervals are sometimes exceedingly wide, assume $T \sim {\rm Tt}(\mu,\, \sigma^2,\, d,\, \mca S_T)$. When the truncation set $\mca S_T$ is bounded, values of $T$ which are close to the endpoints of $\mca S_T$ leads to wide confidence intervals. This behaviour is because there are many values of $\mu$ that would be consistent with the data \citep{lee2016exact, kivaranovic2020length}. On the other hand, when $\mca S_T$ is unbounded, regardless of the values of $T$, the interval length is always bounded. Similar arguments hold for a truncated normal distribution. Figure \ref{fig:ci.length.proj2} displays the lengths of confidence intervals derived using truncated normal and truncated $t$ distributions for a bounded and an unbounded truncation set. The left panel indicates that the length of confidence intervals for both truncated normal and truncated t distributions diverge as the values of $z$ or $t$ approach the edges of the truncation set. This observation certifies the results derived in Theorem \ref{thm:cilength.zttrunc.proj2}. On the other hand, when the truncation set is unbounded, the right panel indicates that the length of confidence intervals for both distributions are bounded. Moreover, the interval length converges to the length of confidence interval obtained from an usual (untruncated) normal or $t$ distributions as the values of the statistics diverge from the edges of the unbounded truncation sets. Additionally, for both cases of bounded and unbounded truncation set, the lengths of confidence intervals for the truncated $t$ distribution are greater or equal than those of the truncated normal distribution.

To measure the quality of the confidence intervals built in \eqref{ztrunc.ci.proj2} and \eqref{ttrunc.ci.proj2}, we inspect their corresponding truncation sets to determine whether they are unbounded. The next theorem verifies that the truncation boundaries provided in Theorems \ref{thm:ztrunc.proj2} and \ref{thm:ttrunc.proj2} are in fact bounded.

\begin{theorem}\label{thm:ciboundary.zttrunc.proj2}
Consider the settings of Theorems \ref{thm:ztrunc.proj2} and \ref{thm:ttrunc.proj2}. Also, let $\big[ \mca V_Z^{-}(\mbf V)\, ,\, \mca V_Z^{+}(\mbf V) \big]$ and $\big[ \mca V_T^{-}(\mbf V,\, W)\, ,\, \mca V_T^{+}(\mbf V,\, W) \big]$ be the truncation sets for the truncated normal and truncated t distributions, derived in \eqref{ztrunc.bounds.proj2} and \eqref{ttrunc.bounds.proj2}, respectively. Then,

\begin{enumerate}[label=(\alph*)]
    \item either one or both truncation boundaries of \eqref{ztrunc.bounds.proj2} are finite, that is,
    \begin{align*}
        -\infty<\mca V_Z^{-} \big( \mbf V \big) \qquad\qquad \trm{or} \qquad\qquad \mca V_Z^{+} \big( \mbf V \big)<\infty.
    \end{align*}
    
    \item either one or both truncation boundaries of \eqref{ttrunc.bounds.proj2} are finite, that is,
    \begin{align*}
        -\infty<\mca V_T^{-} \big( \mbf V,\, W \big) \qquad\qquad \trm{or} \qquad\qquad \mca V_T^{+} \big( \mbf V,\, W \big) <\infty.
    \end{align*}
\end{enumerate}
\end{theorem}
The proof is given in appendix \ref{prf:ciboundary.zttrunc.proj2}. 

\begin{remark}
Theorem \ref{thm:ciboundary.zttrunc.proj2} states that the truncation sets derived in Theorems \ref{thm:ttrunc.proj2} and \ref{thm:cilength.zttrunc.proj2}, for both truncated normal and t distributions,  are bounded. Hence, from Theorem \ref{thm:ciboundary.zttrunc.proj2}, 
\begin{align}
    &E \Big[\, U_Z(Z)- L_Z(Z)\, \Big|\, \big\{ \widehat{\bsy\tau}\, ,\, \widehat{\mbf s} \big\} \Big]= \infty,
    \\[10pt]
    &E \Big[\, U_T(T)- L_T(T)\, \Big|\, \big\{ \widehat{\bsy\tau}\, ,\, \widehat{\mbf s} \big\} \Big]= \infty.
\end{align}
In words, the average lengths of the confidence intervals for $\bsy\eta^T \mbf f$ with known $\sigma^2$, given in \eqref{ztrunc.ci.proj2}, and with unknown $\sigma^2$, given in \eqref{ttrunc.ci.proj2}, are infinite.
\end{remark}

\section{Optimal Post-Detection Inference}
\label{sec:optimal.conditioning.proj2}

Motivated by the fact that post-detection confidence intervals might become extremely wide, we seek  approaches to attain confidence intervals with narrower length properties. Given a target change point, we can essentially think of $\widehat{\bsy\tau}$ as containing two types of change points: those relevant to the discovery of the target change point and those irrelevant to it. The relevant change points are those change points that directly influence the detection of the target change point. In most cases, such points are located around the target change points. Therefore, one solution for improving the performance of post-detection inference is to allow only the relevant change points to participate in the analysis. In other words, the post-detection inference is carried out by only conditioning on a set containing relevant change points. We note that the set which includes these relevant change points is not necessarily a polyhedron. In the following, we will show that by narrowing down the conditional event to the set containing relevant change points, the resulting intervals becomes much shorter. This procedure will also lead to improved power for the associated hypothesis tests.

In the next two sections, we elaborate on two algorithms for conditioning on the relevant change point sets, leading to more powerful post-detection tests and narrower post-detection confidence intervals. We introduce two different setups. Section \ref{sec:global.model.proj2} involves post-detection inference by conditioning on the event that the given target change point is included in the estimated change point set. We refer to this setup as {\it global post-detection} because it verifies the significance of the target change point globally over the entire signal. In contrast, Section \ref{sec:local.model.proj2} involves statistical inference under the condition that the target change point and its adjacent change points are only included in the model. This setup is called local post-detection as it tests the significance of the target change points locally over its adjacent segments.

\subsection{Global Post-Detection Inference}
\label{sec:global.model.proj2}
As explained before, global post-detection assumes that the target change point has been detected by the PRUTF algorithm and information about the rest of the change points is disregarded. In other words, for the given target change point $\widehat{\tau}_j$, we only assume that $\widehat{\tau}_j \in \widehat{\bsy\tau}$. Now, the goal is to carry out post-detection inference by conditioning on the event $\{\,\widehat{\tau}_j\in \widehat{\bsy\tau}\,\}$. Observe that this conditioning is different from conditioning on the entire change point set $\widehat{\bsy\tau}$. We then attempt to compute a post-detection p-value for the hypothesis $H_0:~ \mbf {D}_{\widehat{\tau}_j}\mbf f =0$, as well as building a post-detection confidence interval for $\mbf {D}_{\widehat{\tau}_j}\mbf f$. Depending on whether $\sigma^2$ is known or unknown, the post-detection tests are based on the statistics
\begin{align}\label{zt.glo.proj2}
     Z_j^{\,\trm{glo}}=\frac{\mbf {D}_{\widehat{\tau}_j} \mbf y }{\sigma \|\mbf {D}_{\widehat{\tau}_j}^T\|}\qquad\qquad \trm{and} \qquad\qquad
     T_j^{\,\trm{glo}}=\frac{\mbf {D}_{\widehat{\tau}_j} \mbf y }{\widehat{\sigma}_j^{\,(\trm{glo})} \|\mbf {D}_{\widehat{\tau}_j}^T\|}
\end{align}
which, under the null hypothesis, are distributed as the standard truncated normal and truncated $t$ distributions, respectively. Given $\{ \widehat{\tau}_j \in \widehat{\bsy\tau} \}$, we split $\mbf y$ into two subvectors $\mbf y_{1,\, j}^{\,\trm{glo}}= (y_1, \ldots, y_{\widehat{\tau}_j})$ and $\mbf y_{2,\, j}^{\,\trm{glo}}= (y_{\widehat{\tau}_j+1}, \ldots, y_n)$. The projection matrices $\mbf P_{k,\, j}^{\,\trm{glo}}$ associated with the subvectors $\mbf y_{k,\, j}^{\trm{glo}}$, $k=1,2$, the block diagonal matrix $\mbf P_{ j}^{\trm{glo}}= \diag{\mbf P_{1,\, j}^{\trm{glo}}\, ,\, \mbf P_{2,\, j}^{\trm{glo}}}$ and the sufficient statistic $\mbf V_j^{\trm{glo}}$  can be defined accordingly (see Section \ref{sec:post.detection.proj2}). The next theorem explains how to specify the truncation sets for the underlying distributions. We provide the proof of the theorem in Appendix \ref{prf:global.proj2}.

\begin{theorem}\label{thm:global.proj2}
Consider the statistics $  Z_j^{\,\trm{glo}}$ and $ T_j^{\,\trm{glo}}$ as defined in \eqref{zt.glo.proj2}.

\begin{enumerate}[label=(\alph*)]
    \item The conditional distribution of $ Z_j^{\,\trm{glo}}$ given $\{\widehat{\tau}_j\in \widehat{\bsy\tau}\}$, under the null hypothesis $H_0:~ \mbf {D}_{\widehat{\tau}_j}\mbf f =0$, is the standard normal truncated to the set $\big( -\infty\, ,\, \mca V_{Z_j}^{-(\trm{glo})} \big]\, \bigcup\,\, \big[ \mca V_{Z_j}^{+(\trm{glo})}\, ,\, \infty \big)$, where
    \begin{align}\label{glo.zbounds.proj2}
        \mca V_{Z_j}^{-(glo)}&= \frac{-\lambda_j\, \mbf D_{\widehat{\tau}_j}\, \mbf D_{\widehat{\tau}_j}^T + \mbf D_{\widehat{\tau}_j}\, \mbf D_{-\widehat{\tau}_j}^T\, \widehat{\mbf u}_{\lambda_j,\, -\widehat{\tau}_j} }{ \sigma\,  \|\mbf D_{\widehat{\tau}_j}^T \|},
        \\[10pt]
        \mca V_{Z_j}^{+(glo)}&= \frac{\lambda_j\, \mbf D_{\widehat{\tau}_j}\, \mbf D_{\widehat{\tau}_j}^T + \mbf D_{\widehat{\tau}_j}\, \mbf D_{-\widehat{\tau}_j}^T\, \widehat{\mbf u}_{\lambda_j,\, -\widehat{\tau}_j} }{ \sigma\,  \|\mbf D_{\widehat{\tau}_j}^T \|}.
    \end{align}
 
    \item The conditional distribution of $T_j^{\,\trm{glo}}$ given $\{\widehat{\tau}_j\in \widehat{\bsy\tau}\}$, under the null hypothesis, is $t$ distribution with $d^{\,(\trm{glo})}=n-2$ degrees of freedom and truncated to the set \linebreak
    $\big( -\infty\, ,\, \mca V_{T_j}^{-(\trm{glo})} \big]\, \bigcup\,\, \big[ \mca V_{T_j}^{+(\trm{glo})}\, ,\, \infty \big)$, where
    \begin{align}\label{glo.tbounds.proj2}
        \mca V_{T_j}^{-(glo)}&= \frac{-\lambda_j\, \mbf D_{\widehat{\tau}_j}\, \mbf D_{\widehat{\tau}_j}^T + \mbf D_{\widehat{\tau}_j}\, \mbf D_{-\widehat{\tau}_j}^T\, \widehat{\mbf u}_{\lambda_j,\, -\widehat{\tau}_j} }{ \widehat{\sigma}^{\,(\trm{glo})} \, \|\mbf D_{\widehat{\tau}_j}^T \|},
        \\[10pt]
        \mca V_{T_j}^{+(glo)}&= \frac{\lambda_j\, \mbf D_{\widehat{\tau}_j}\, \mbf D_{\widehat{\tau}_j}^T + \mbf D_{\widehat{\tau}_j}\, \mbf D_{-\widehat{\tau}_j}^T\, \widehat{\mbf u}_{\lambda_j,\, -\widehat{\tau}_j} }{ \widehat{\sigma}^{\,(\trm{glo})} \,  \|\mbf D_{\widehat{\tau}_j}^T \|},
    \end{align}
    and $\widehat{\sigma}_j^{\,2 \,(\trm{glo})}= \big\| \big(\mbf I-\mbf P_{_j}^{\trm{glo}} \big)\, \mbf y \big\|^2/ d^{\,(\trm{glo})}$.
\end{enumerate}
\end{theorem}

We note that the TN and Tt statistics, provided in \eqref{tn.stat.proj2} and \eqref{tt.stat.proj2}, respectively, can be constructed using the corresponding distributions in both parts (a) and (b) of Theorem \ref{thm:global.proj2}. Therefore, We can apply the same procedure as in Section \ref{sec:post.detection.proj2} to compute post-detection p-values and confidence intervals.

It is  worth mentioning that  $\widehat{\sigma}_j^{\,2 \,(\trm{glo})}$ may not be a suitable estimator of $\sigma^2$. This is because global post-detection only assumes that $\widehat{\tau}_j \, \in \, \widehat{\bsy\tau}$. Therefore, in the case that there exit other change points in $\widehat{\bsy\tau}$, $\widehat{\sigma}_j^{\,2 \,(\trm{glo})}$ is unable to accurately estimate the variation in the observations. As an alternative, we can apply Median Absolute Deviation (MAD) proposed by \cite{hampel1974influence}, to robustly estimate $\sigma^2$. Also, see \cite{baranowski2019narrowest} for more details about MAD estimation.

Let $[L_{Z_j}^{\trm{glo}}\, ,\, U_{Z_j}^{\trm{glo}}]$ and $[L_{T_j}^{\trm{glo}}\, ,\, U_{T_j}^{\trm{glo}}]$ be confidence intervals derived in the same manner as \eqref{ztrunc.ci.proj2} and \eqref{ttrunc.ci.proj2} using the pivotal quantities $Z_j^{\,\trm{glo}}$ and $T_j^{\,\trm{glo}}$. As the lower and upper bounds associated with their distributions are unbounded, we suspect that the underlying confidence intervals have finite expected lengths. In such a case, \cite{kivaranovic2020length} have provided an upper bound for the confidence intervals derived from a truncated normal distribution. In the following theorem, we will also give such an upper bound for confidence intervals using a truncated $t$ distribution.

\begin{theorem}\label{thm:upperbound.zttrunc.global.proj2}
Let $\big[L_{Z_j}^{\trm{glo}}\,,\, U_{Z_j}^{\trm{glo}}\big]$ and $\big[L_{T_j}^{\trm{glo}}\,,\, U_{T_j}^{\trm{glo}}\big]$ be  $(1-\alpha)\%$ confidence intervals derived from truncated normal and truncated $t$ distributions given in Theorem \ref{thm:global.proj2}.
\begin{enumerate}[label=(\alph*)]
    \item The length of confidence intervals derived from the truncated normal distribution is always upper bounded by
    \begin{align}\label{z.upper.bound.proj2}
    U_{Z_j}^{\trm{glo}}-L_{Z_j}^{\trm{glo}}& ~ \overset{a.s.}{\leq} ~ 2\, \sigma\, \Phi^{-1}\left(1-\frac{\alpha}{2}\right) ~+~ \mca V_{Z_j}^{\,+(\trm{glo})}-\mca V_{Z_j}^{\,-(\trm{glo})} \,.
    \end{align}
    
    \item For $d^{\,(\trm{glo})} \geq 3$ and under the condition
    \begin{align}\label{t.upper.bound.condition.proj2}
    \alpha\,\, G_{d^{\,(\trm{glo})}} \left( -\, \frac{\mca V_{T_j}^{+(\trm{glo})}-\mca V_{T_j}^{-(\trm{glo})} }{ 2\, \widehat{\sigma}_j^{\, (\trm{glo})}}\right) \,\geq\,G_{d^{\,(\trm{glo})}} \left( G_{d^{\,(\trm{glo})}}^{-1} \left( \frac{\alpha}{2} \right)- \frac{ \mca V_{T_j}^{+(\trm{glo})}-\mca V_{T_j}^{-(\trm{glo})} }{ 2\, \widehat{\sigma}_j^{\, (\trm{glo})}} \right),
    \end{align}
    the length of confidence intervals derived from the truncated $t$ distribution is upper bounded by
    \begin{align}\label{t.upper.bound.proj2}
    U_{T_j}^{\trm{glo}}-L_{T_j}^{\trm{glo}}& ~ \stackrel{a.s.}{\leq} ~ 2\, \widehat{\sigma}_j^{\, (\trm{glo})}\, G_{d^{\,(\trm{glo})}}^{-1} \left(1-\frac{\alpha}{2}\right) ~+~ \mca V_{T_j}^{+(\trm{glo})}-\mca V_{T_j}^{-(\trm{glo})} \,.
    \end{align}
    
\end{enumerate}
\end{theorem}
A proof is provided in Appendix \ref{prf:upperbound.ttrunc.proj2}.


\subsection{Local Post-Detection Inference}
\label{sec:local.model.proj2}
Unlike in Section \ref{sec:global.model.proj2}, here we are dealing with situations where the most relevant estimated change points to the target change point are involved in our inferential analysis.
Obviously, the selection of $\widehat{\tau}_j$ as a change point relies on other estimated change points. Thus, conditioning only on $\{\widehat{\tau}_j\in \widehat{\bsy\tau}\}$ is insufficient to decide whether $\widehat{\tau}_j$ is a meaningful change location. This fact indicates the need for knowledge about other change points. However, as previously discussed, conditioning our inference on the entire change point set yields an undesirable output such as wide confidence intervals. Note that the local nature of change point setting \citep{niu2012screening} dictates that inferences for a change point should depend on its adjacent change points. For instance, only the immediate neighboring change points $\widehat{\tau}_{j-1}$ and $\widehat{\tau}_{j+1}$ play a role in the detection of the target change point $\widehat{\tau}_j$, and the rest remain irrelevant. This fact has motivated us to develop the local post-detection algorithm.

Following the local property of change points, we suggest an idea for post-detection inference which leads to higher-powered tests and shorter confidence intervals. The idea is to condition the post-detection inference on the target change point as well as on its adjacent change points. In particular, for the given target change point $\widehat{\tau}_j$, the goal is to test the hypothesis $H_0:~ \mbf D_{\widehat{\tau}_j}\,\mbf f=0$, given $\big\{ \big(\widehat{\tau}_{j-1},\, \widehat{\tau}_{j},\, \widehat{\tau}_{j+1} \big) \in \widehat{\bsy\tau} \big\}$. In fact, this conditioning creates a new change point problem over the shorter subsignal $\big( f_{\widehat{\tau}_{j-1}+1},\, \ldots,\, f_{\widehat{\tau}_{j+1}} \big)$. We first define some notations for ease of exposition. 
Let $\mbf y^{\trm{loc}}_j= \big( \mbf y_{j-1}\, ,\, \mbf y_j \big)= \big( y_{\widehat{\tau}_{j-1}+1}\, ,\, \ldots\, ,\, y_{\widehat{\tau}_{j+1}} \big)$ and $\mbf f^{\trm{loc}}_j=\big( \mbf f_{j-1}\, ,\, \mbf f_j \big)= \big( f_{\widehat{\tau}_{j-1}+1}\, ,\, \ldots\, ,\, f_{\widehat{\tau}_{j+1}} \big)$ denote the subvectors of $\mbf y$ and $\mbf f$ from $\widehat{\tau}_{j-1}+1$ to $\widehat{\tau}_{j+1}$. Also, let $\mbf D_j$ be the $\big( \widehat{\tau}_{j+1}-\widehat{\tau}_{j-1}-r-1 \big) \times \big( \widehat{\tau}_{j+1}-\widehat{\tau}_{j-1} \big)$ version of matrix $\mbf D$. Applying these notations, the hypothesis of interest can be re-expressed as $H_0:~ \mbf D_{j,\,\Delta_j}\,\mbf f^{\trm{loc}}_j=0$, where $\Delta_j=\widehat{\tau}_j-\widehat{\tau}_{j-1}$ and $\mbf D_{j,\,\Delta_j}$ is the $\Delta_j$-th row of $\mbf D_{j}$. For inference, similar approach taken in the global post-detection can be applied, but this time using $\mbf y^{\trm{loc}}_j$ and $\mbf D_j$. Depending on whether $\sigma^2$ is known or unknown, the post-detection tests are based on statistics
\begin{align}\label{zt.loc.proj2}
     Z_j^{\,\trm{loc}}=\frac{\mbf D_{j,\,\Delta_j}\, \mbf y^{\trm{loc}}_j}{\sigma\, \big\|\mbf D_{j,\,\Delta_j} \big\|}
     \qquad\qquad,\qquad\qquad
     T_j^{\,\trm{loc}}=\frac{\mbf D_{j,\,\Delta_j}\, \mbf y^{\trm{loc}}_j}{ \widehat{\sigma}^{\, (\trm{loc})}_j \big\| \mbf D_{j,\,\Delta_j} \big\|}.
\end{align}
We will obtain the distributions of these statistics in the next theorem whose proof is similar to that of Theorem \ref{thm:global.proj2} by replacing $\mbf y,\, \mbf D,\, \widehat{\tau}_j$, $\mbf P_j^{\trm{\,glo}}$ and $\widehat{\sigma}^{\,2 \,(\trm{glo})}_j$ with $\mbf y^{\trm{loc}}_j,\, \mbf D_j,\, \Delta_j$, $\mbf P_{j}^{\trm{loc}}$ and $\widehat{\sigma}^{\,2 \,(\trm{loc})}_j$.

\begin{theorem}\label{thm:local.proj2}
Consider the statistics $Z_j^{\,{\rm loc}}$ and $T_j^{\, {\rm loc}}$ defined in \eqref{zt.loc.proj2}.

\begin{enumerate}[label=(\alph*)]
    \item Given $\big\{ \big( \widehat{\tau}_{j-1}\, ,\, \widehat{\tau}_{j},\, \widehat{\tau}_{j+1} \big) \in \widehat{\bsy\tau} \big\}$, the conditional distribution of $ Z_j^{\,{\rm loc}}$, under the null hypothesis $H_0:~ \mbf {D}_{\widehat{\tau}_j}\mbf f =0$, or equivalently $H_0:~ \mbf D_{j,\,\Delta_j}\,\mbf f^{{\rm loc}}_j=0$,  is the standard normal truncated to $\big( -\infty\, ,\, \mca V_{Z_j}^{-({\rm loc})} \big]\, \bigcup\,\, \big[ \mca V_{Z_j}^{+({\rm loc})}\, ,\, \infty \big)$, where
    \begin{align}\label{loc.zbounds.proj2}
        \mca V_{Z_j}^{-({\rm loc})}&= \frac{-\lambda_j\, \mbf D_{j,\, \Delta_j}\, \mbf D_{j,\, \Delta_j}^T + \mbf D_{j,\, \Delta_j}\, \mbf D_{j,\, -\Delta_j}^T\, \widehat{\mbf u}_{\lambda_j,\, -\widehat{\tau}_j} }{ \sigma \, \big\| \mbf D_{j,\, \Delta_j}^T \big\|},
        \\[10pt]
        \mca V_{Z_j}^{+({\rm loc})}&= \frac{\lambda_j\, \mbf D_{j,\, \Delta_j}\, \mbf D_{j,\, \Delta_j}^T + \mbf D_{j,\, \Delta_j}\, \mbf D_{j,\, -\Delta_j}^T\, \widehat{\mbf u}_{\lambda_j,\, -\widehat{\tau}_j} }{ \sigma \, \big\| \mbf D_{j,\, \Delta_j}^T \big\|}.
    \end{align}

    \item Given $\big\{ \big( \widehat{\tau}_{j-1}\, ,\, \widehat{\tau}_{j}\, ,\, \widehat{\tau}_{j+1} \big) \in \widehat{\bsy\tau} \big\}$, the conditional distribution of $T_j^{{\rm loc}}$, under the null hypothesis, is $t$-distribution with $d_j^{\,({\rm loc})}=(\widehat{\tau}_{j+1}-\widehat{\tau}_{j-1})-2$ truncated to the set $\big( -\infty\, ,\, \mca V_{T_j}^{-({\rm loc})} \big]$\, $\bigcup\,\, \big[\mca V_{T_j}^{+({\rm loc})}\, ,\, \infty \big)$, where
        \begin{align}\label{loc.tbounds.proj2}
        \mca V_{T_j}^{-({\rm loc})}&=  \frac{-\lambda_j\, \mbf D_{j,\, \Delta_j}\, \mbf D_{j,\, \Delta_j}^T + \mbf D_{j,\, \Delta_j}\, \mbf D_{j,\, -\Delta_j}^T\, \widehat{\mbf u}_{\lambda_j,\, -\widehat{\tau}_j} }{ \widehat{\sigma}^{\,(\trm{loc})}_j\, \|\mbf D_{j,\, \Delta_j}^T \|},
        \\[10pt]
        \mca V_{T_j}^{+({\rm loc})}&= \frac{\lambda_j\, \mbf D_{j,\, \Delta_j}\, \mbf D_{j,\, \Delta_j}^T + \mbf D_{j,\, \Delta_j}\, \mbf D_{j,\, -\Delta_j}^T\, \widehat{\mbf u}_{\lambda_j,\, -\widehat{\tau}_j} }{ \widehat{\sigma}^{\,(\trm{loc})}_j\, \|\mbf D_{j,\, \Delta_j}^T \|},
    \end{align}
    and $\widehat{\sigma}^{\,2 \,({\rm loc})}_j=\, \big\|(\mbf I-\mbf P_j^{{\rm loc}})\,\mbf y_j^{{\rm loc}} \big\|^2/d_j^{\, ({\rm loc})}$.
\end{enumerate}
\end{theorem}
In the same manner as described in Section \ref{sec:post.detection.proj2}, the TN and Tt statistics can be employed to perform hypothesis testings and construct confidence intervals.

\begin{remark}\label{lem:length.upbound.locloc.proj2}
From Theorem \ref{thm:local.proj2}, it turns out that the lower and upper limits of truncation sets using local post-detection inference are infinite. Therefore, similar to global post-detection inference, the expected lengths of confidence intervals associated with such distributions are upper bounded. In particular, let $\big[ L_{Z_j}^{{\rm loc}}\,,\, U_{Z_j}^{{\rm loc}} \big]$ and $\big[ L_{T_j}^{{\rm loc}}\,,\, U_{T_j}^{{\rm loc}} \big]$ be  $(1-\alpha)\%$ confidence intervals derived from truncated normal and truncated $t$ distributions, respectively, provided in Theorem \ref{thm:local.proj2}. Hence, the lengths of such intervals are upper bounded by

\begin{itemize}
    \item when $\sigma^2$ is known, the length of confidence intervals is always upper bounded by
    \begin{align}\label{zlocal.upper.bound.proj2}
    U_{Z_j}^{{\rm loc}}-L_{Z_j}^{{\rm loc}}& ~\overset{a.s.}{\leq}~ 2\,\sigma\,\Phi^{-1}\left(1-\frac{\alpha}{2}\right)+ \mca V_{Z_j}^{+({\rm loc})}-\mca V_{Z_j}^{-({\rm loc})} \,.
    \end{align}
    
    \item when $\sigma^2$ is unknown, for $d^{\,(\trm{loc})} \geq 3$ and under the condition
    \begin{align}\label{tlocal.upper.bound.condition.proj2}
    \alpha\,\, G_{d^{\,(\trm{loc})}} \left( -\, \frac{\mca V_{T_j}^{\,+(\trm{loc})}-\mca V_{T_j}^{\,-(\trm{loc})} }{ 2\, \widehat{\sigma}_j^{\, (\trm{loc})}}\right) \,\geq\,G_{d^{\,(\trm{loc})}} \left( G_{d^{\,(\trm{loc})}}^{-1} \left( \frac{\alpha}{2} \right)- \frac{ \mca V_{T_j}^{\,+(\trm{loc})}-\mca V_{T_j}^{\,-(\trm{loc})} }{ 2\, \widehat{\sigma}_j^{\, (\trm{loc})}} \right),
    \end{align}
    the length of confidence intervals is upper bounded by
    \begin{align}\label{tlocal.upper.bound.proj2}
    U_{T_j}^{{\rm loc}}-L_{T_j}^{{\rm loc}}& ~\overset{a.s.}{\leq}~ 2\, \widehat{\sigma}_j^{\,({\rm loc})}\,G_{d^{\,({\rm loc})}}^{-1}\left(1-\frac{\alpha}{2}\right)+ \mca V_{T_j}^{\,+({\rm loc})}-\mca V_{T_j}^{\,-({\rm loc})} \,.
    \end{align}
\end{itemize}
\end{remark}

\section{Numerical Studies}
\label{sec:simulations}
In this section, we investigate the performance of our proposed post-detection inference approaches for a piecewise constant signal ($r=0$) and a piecewise linear signal $(r=1)$. We compare the performance of the approaches in terms of the empirical power for hypothesis testings and coverage probabilities for confidence intervals. We also apply the implementation of the proposed methods to three real datasets, which have been used to estimate change points using the PRUTF algorithm in \cite{mehrizi2020detection}.

\subsection{Simulations}

In order to investigate our proposed approaches for post-detection inference, we consider two signals: piecewise constant and piecewise linear. Suppose that $\mbf f$ is a piecewise constant or linear signal of size $n=500$, with four change points, $J_0=4$, at locations $\bsy\tau= \big\{\, 100\, ,\, 200\, ,\, 300\, ,\, 400\, \big\}$. For the piecewise constant signal, we consider a signal with the starting point 0 and the jump sizes $\delta$. That is,
\begin{align*}
    \hspace{-2cm} f_t^{\,\trm{con}}= \left\{
    \begin{array}{llllllllllll}
        0 &&& \trm{for} && 1 \leq t \leq 100 & \trm{or} & 201 \leq t \leq 300 & \trm{or} & 401 \leq t \leq 500, \\[8pt]
        \delta &&& \trm{for} && 101 \leq t \leq 200 & \trm{or} & 301 \leq t \leq 400\,,
    \end{array}
    \right.
\end{align*}
where $\delta \in \big\{\, 1\, ,\, 1.5\, ,\, 2\, ,\, \ldots,\, 4.5\, ,\, 5\, \big\}$. Also, for the piecewise linear signal, we consider \begin{align*}
    f_t^{\,\trm{lin}}= \left\{
    \begin{array}{llllllllllll}
        \delta \, ( -0.5 +  t) &&& \trm{for} && 1 \leq t \leq 100 & \trm{or} & 201 \leq t \leq 300& \trm{or} & 401 \leq t \leq 500,
        \\[8pt]
        \delta \, ( 0.5 -  t) &&& \trm{for} && 101 \leq t \leq 200 & \trm{or} & 301 \leq t \leq 400\,.
    \end{array}
    \right.
\end{align*}

\begin{figure}[!b]
\begin{center}
\begin{subfigure}{.4\textwidth}
  \centering
  \includegraphics[width=1\linewidth]{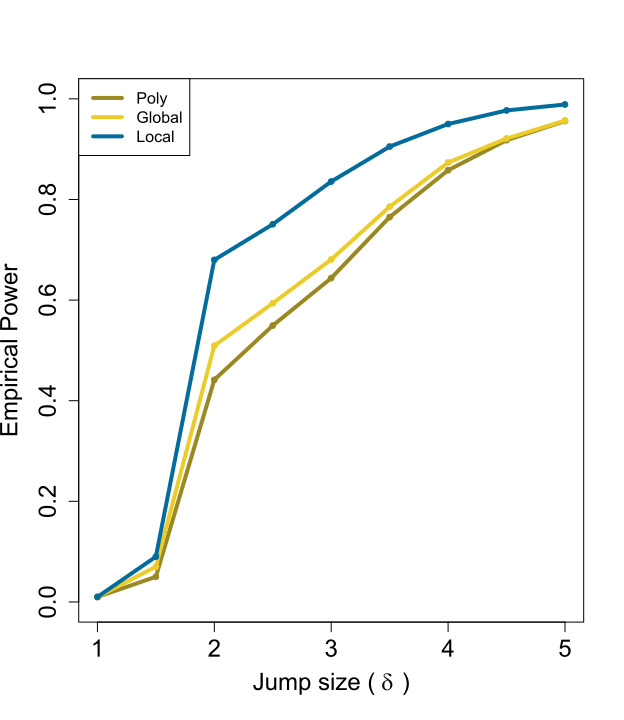}
  \caption{Piecewise constant with known $\sigma^2$.}
\end{subfigure}
\begin{subfigure}{.4\textwidth}
  \centering
  \includegraphics[width=1\linewidth]{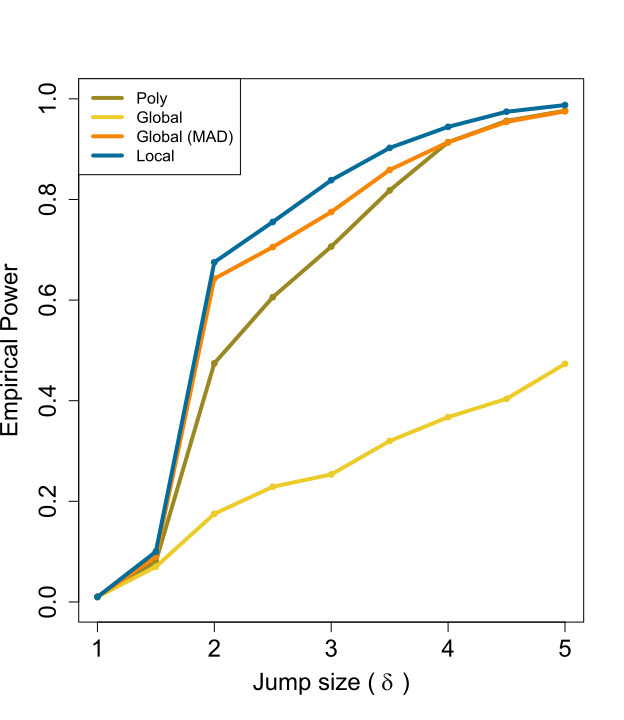}
  \caption{Piecewise constant with unknown $\sigma^2$.}
\end{subfigure}
\begin{subfigure}{.4\textwidth}
  \centering
  \includegraphics[width=1\linewidth]{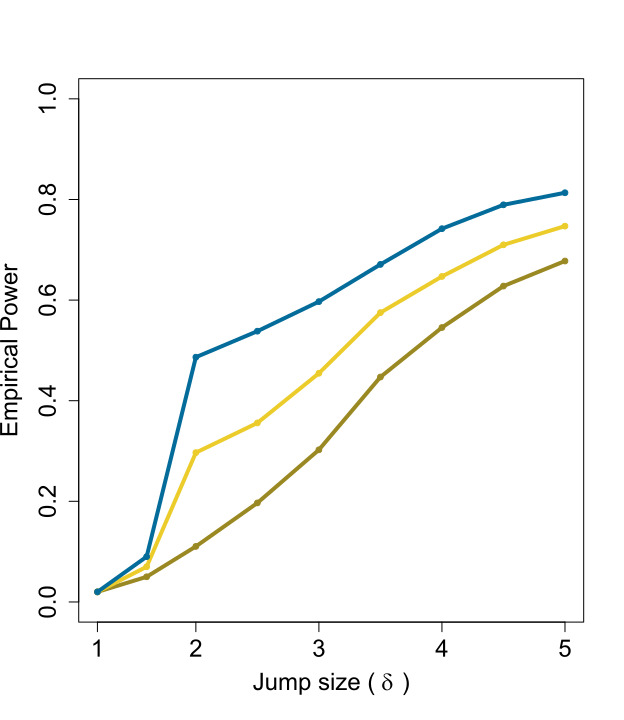}
  \caption{Piecewise linear with known $\sigma^2$.}
\end{subfigure}
\begin{subfigure}{.4\textwidth}
  \centering
  \includegraphics[width=1\linewidth]{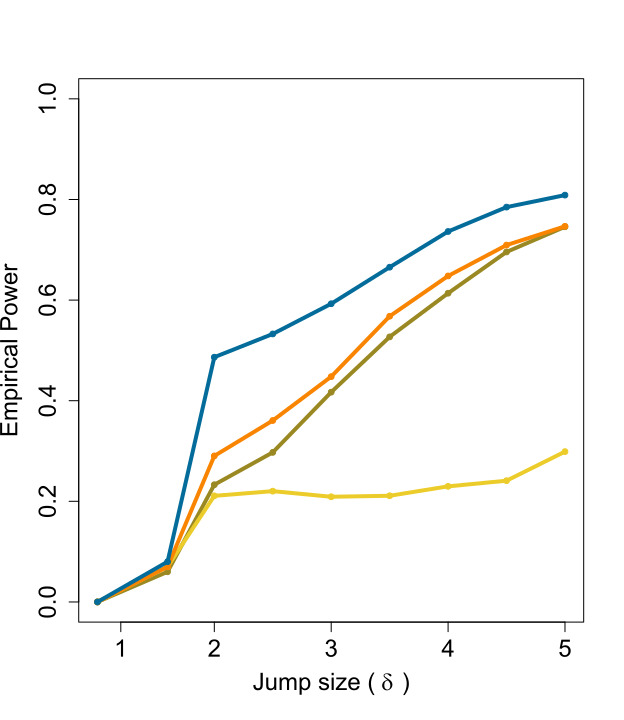}
  \caption{Piecewise linear with unknown $\sigma^2$.}
\end{subfigure}
\caption{ The empirical powers of the three proposed approaches, polyhedron, global and local post-detection inference. The results are provided for both cases when $\sigma^2$ is assumed known and unknown. The two top panels display the empirical powers for the piecewise constant signal. Also, the two bottom panels show the empirical powers for the piecewise linear signal. The solid orange lines in both panels (b) and (d) display the empirical powers of the global post-detection approach when $\widehat{\sigma}_j^{2\,(\trm{glo})}$, an estimation of $\sigma^2$, is replaced with MAD estimation.}
\label{fig:pdi_power}
\end{center}
\end{figure}

We have generated a sample $\mbf y$ from $N\big( \mbf f, \sigma^2\, \mbf I \big)$, with $\sigma^2= 1$, for both piecewise constant and piecewise linear signals, and implemented PRUTF to the estimate change points. Next, we have applied our proposed approaches: polyhedron post-detection ({\it poly}), global post-detection ({\it global}) and local post-detection ({\it local}) inference procedures. Over $N=10000$ repetitions, we have computed the empirical power of a change point hypothesis using the three methods {\it poly, global} and {\it local}, for both cases of known and unknown $\sigma^2$. Recall that the empirical power is defined as the ratio between the number of times the true change point is correctly detected and has $p$-value less than $\alpha$, and the number of times the true change point is correctly detected, see \cite{hyun2018post}. Here, $\alpha$ is set to 0.05. The results of the second change point, $\tau_2= 200$, are displayed in Figure \ref{fig:pdi_power}. For the piecewise constant signal, the empirical powers are computed over repetitions that contain the underlying change point. This is the spike contrast put forward by \cite{hyun2018exact}. Whereas the window contrast, used in \cite{jewell2019testing}, with size, $h=15$ is employed for the computation in the piecewise linear signal. This is because PRUTF mostly estimates change points near but not exactly at change points when the polyhedron degree $r$ increases.

Panels (a) and (b) of Figure \ref{fig:pdi_power} provide the empirical powers for the piecewise constant signal with both known and unknown $\sigma^2$. When $\sigma^2$ is known, the local and global post-detection approaches outperform the polyhedron post-detection approach, as expected. In the case of unknown $\sigma^2$, the local post-detection approach again performs better than the polyhedron one. However, the global post-detection approach has a poor performance due to inaccurate $\sigma^2$ estimation, see Section \ref{sec:global.model.proj2}. By replacing $\widehat{\sigma}_j^{2\,(\trm{glo})}$ in Theorem \ref{thm:global.proj2} with the MAD estimation, the empirical power of the global post-detection approach has improved. The results for the piecewise linear signal are also plotted in panels (c) and (d) of Figure \ref{fig:pdi_power}. Similar patterns are observed for this signal, as well. 

Moreover, we have computed the coverage probability for confidence intervals, obtained using the polyhedron, global and local post-detection methods. The outputs are reported in Table \ref{tab:covprob}.
\begin{table}[!hb]
    \begin{center}
    {\footnotesize
    \begin{tabular}{ cc cc cccccc cc cccccccc}
        \cline{1-18}
         \parbox[t]{3mm}{\multirow{6}{*}{\rotatebox[origin=c]{90}{{\scriptsize Piecewise Constant}}}}
          && && \multicolumn{6}{c}{Known $\sigma^2$} &&& \multicolumn{6}{c}{Unknown $\sigma^2$} \\
         \cline{5-9} \cline{12-18}
        && $\delta$ && Poly && Global && Local &&& Poly && Global && Global (MAD) && Local \\
         \cline{3-3} \cline{5-5}
         \cline{7-7} \cline{9-9} \cline{12-12} \cline{14-14} \cline{16-16} \cline{18-18}
          && 2   && 0.9515 && 0.9527 && 0.9515 &&& 0.9515 && 0.9708 && 0.9527 && 0.9504 \\ 
          && 3   && 0.9554 && 0.9564 && 0.9554 &&& 0.9554 && 0.9817 && 0.9577 && 0.9564 \\ 
          && 4   && 0.9547 && 0.9586 && 0.9547 &&& 0.9605 && 0.9874 && 0.9601 && 0.9551 \\ 
          && 5   && 0.9543 && 0.9531 && 0.9543 &&& 0.9618 && 0.9889 && 0.9553 && 0.9545 \\        \cline{2-18}
         \parbox[t]{3mm}{\multirow{6}{*}{\rotatebox[origin=c]{90}{{\scriptsize Piecewise Linear}}}}
          && && \multicolumn{6}{c}{Known $\sigma^2$} &&& \multicolumn{6}{c}{Unknown $\sigma^2$} \\
         \cline{5-9} \cline{12-18}
       && $\delta$ && Poly && Global && Local &&& Poly && Global && Global (MAD) && Local \\
         \cline{3-3} \cline{5-5}
         \cline{7-7} \cline{9-9} \cline{12-12} \cline{14-14} \cline{16-16} \cline{18-18}
         && 2 && 0.9473 && 0.9543 && 0.9473 &&& 0.9473 && 0.9660 && 0.9450 && 0.9473 \\ 
         && 3 && 0.9550 && 0.9555 && 0.9550 &&& 0.9570 && 0.9737 && 0.9581 && 0.9555 \\ 
         && 4 && 0.9453 && 0.9491 && 0.9453 &&& 0.9491 && 0.9779 && 0.9514 && 0.9441 \\ 
         && 5 && 0.9431 && 0.9419 && 0.9431 &&& 0.9506 && 0.9786 && 0.9426 && 0.9431 \\ 

        \cline{1-18}
    \end{tabular}
    \caption{The coverage probabilities of confidence intervals obtained using polyhedron, global and local post-detection approaches. The results are reported for four values of $\delta \in \{\, 2\, ,\, 3\, ,\, 4\, ,\, 5\, \}$. 
    }
    \label{tab:covprob}}
    \end{center}
\end{table}

\subsection{Real Data Analysis}
For real datasets, \cite{mehrizi2020detection} have analyzed three datasets: UK House Price Index,  the GISS Surface Temperature and COVID-19. In the following, we will apply our proposed post-detection inference approaches to evaluate the significance of change points estimated using PRUTF for these datasets.

\begin{figure}[!t]
\begin{center}
\begin{subfigure}{.47\textwidth}
  \centering
  \includegraphics[width=1\linewidth]{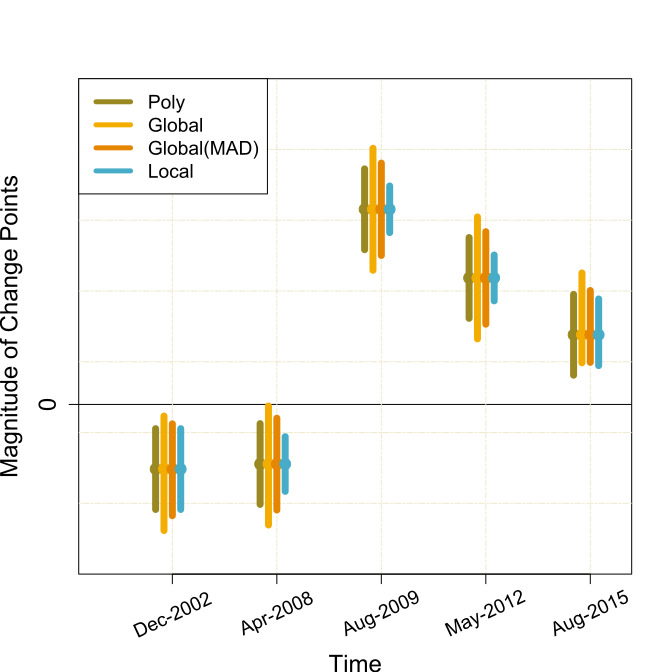}
  \caption{CIs For HPI Data.}
\end{subfigure}
\begin{subfigure}{.47\textwidth}
  \centering
  \includegraphics[width=1\linewidth]{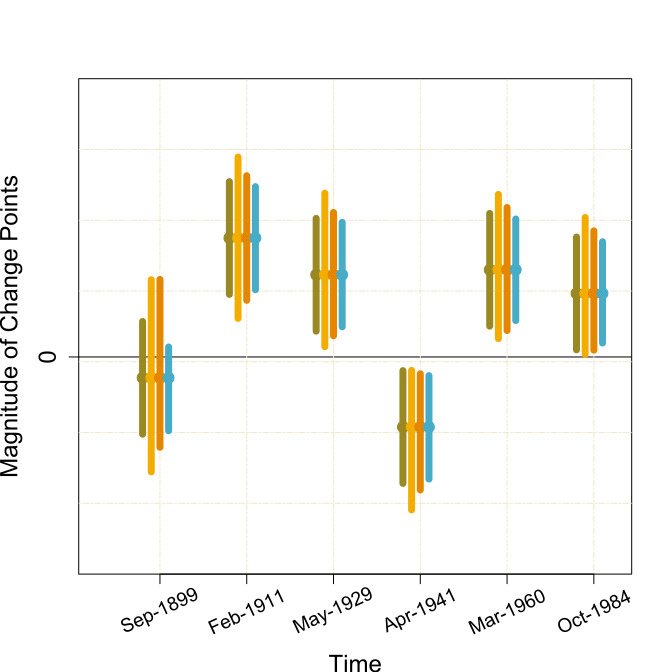}
  \caption{CIs For GISTEMP Data.}
\end{subfigure}
\caption [Post-Detection Confidence Intervals For UK HPI and GISTEMP Datasets] {Valid post-detection confidence intervals for both the UK HPI dataset (left panel) and GISTEMP dataset (right panel). For each estimated change point, valid $\% \, 95$ post-detection confidence intervals, obtained from polyhedron, global using both pooled and MAD variances, and local post-detection inference approaches, are displayed. }
\label{fig:pdi_hpi_gistemp}
\end{center}
\end{figure}

\begin{example}\label{HPIdata.example}

The UK House Price Index (HPI) is a National Statistic that shows changes in the value of residential properties in England, Scotland, Wales and Northern Ireland. The HPI measures the price changes of residential housing by calculating the price of completed house sale transactions as a percentage change from some specific start date. The UK HPI uses the hedonic regression model as a statistical approach to producing estimates of the change in house prices for each period. For more details, see \url{https://landregistry.data.gov.uk/app/ukhpi}.
Many researchers, including \cite{baranowski2019narrowest} and \cite{anastasiou2019detecting}, have studied the UK HPI dataset by carrying out change point analysis.
In the current study, we consider monthly percentage changes in the UK HPI at Tower Hamlets (an east borough of London) from January 1996 to November 2018.

The PRUTF algorithm has found five change points located at months: December 2002, April 2008 and August 2009, May 2012  and August 2015. See \cite{mehrizi2020detection} for more details. We have provided the results of post-detection inference for these change points in panel (a) of Figure \ref{fig:pdi_hpi_gistemp}. The plot displays $\% \, 95$ post-detection confidence intervals for the change points. Based on the polyhedron and local post-detection results, we have concluded that all five estimated change points are significant as their corresponding intervals excluded zero. 


\end{example}


\begin{example}\label{GISTEMPdata.example}

The Goddard Institute for Space Studies (GISS) monitors broad global changes around the world. The GISS Surface Temperature Analysis (GISTEMP) is an estimate of the global surface temperature changes (see \url{https://www.giss.nasa.gov}). In the analysis of GISTEMP data, the temperature anomalies are used rather than the actual temperatures. A temperature anomaly is a difference from an average or baseline temperature. The baseline temperature is typically computed by averaging thirty or more years of temperature data (1951 to 1980 in the current dataset). A positive anomaly indicates the observed temperature was warmer than the baseline, while a negative anomaly indicates the observed temperature was cooler than the baseline. For more details see \cite{lenssen2019improvements}.

The GISTEMP dataset has been frequently explored in change point literature; for example, see \cite{james2015change} and \cite{baranowski2019narrowest}.
We have considered the monthly land-ocean temperature anomalies recorded from January 1880 to August 2019 (see \url{https://data.giss.nasa.gov/gistemp}). For this dataset, \cite{mehrizi2020detection} have identified six change points located at months: September 1899, February 1911, May 1929, April 1941, March 1960, October 1984.

We have applied our proposed post-detection inference approaches to assess the significance of these estimated change points by computing the associated confidence intervals. The results are provided in the right panel of Figure \ref{fig:pdi_hpi_gistemp}. The outputs obtained from both polyhedron and local post-detection inference have confirmed the significance of all estimated change points except the one in September 1899.


\end{example}


\begin{example}
Since the initial outbreak of Coronavirus Disease (COVID-19) in Wuhan, China, in mid-November 2019, the virus has rapidly spread throughout the world. The pandemic has infected 167.25 million people and killed more than 3.46 (up to March 4, 2021) \url{https://covid19.who.int/}, greatly stressing public health systems and adversely influencing global society and economies. Therefore, every country has attempted to slow down the transmission rate by various regional and national policies such as the declaration of national emergencies, quarantines and mass testing. Of vital interest to governments is understanding the pattern of the epidemic growth and assessing the effectiveness of policies undertaken. A scientist can investigate these matters by analyzing the sequence of infection data for COVID-19. Change point detection is one possible framework for studying the behaviour of COVID-19 infection curves and retrospectively investigating the effectiveness of interventions. By detecting the locations of alterations in the curves, change point analysis gives us insights into changes in the transmission rate or efficiency of interventions. It also enables us to raise warning signals if the disease pattern changes.

We have analyzed the logarithm of the cumulative daily number of confirmed cases using a piecewise linear model. The choice of the piecewise linear model is natural because the slope of each segment, estimated using the detected change points, indicates the growth rate of the coronavirus. Consequently, these slopes allow us to compare the virus growth rate among estimated segments and evaluate the effectiveness of undertaken strategies. Additionally, the linear trend of the last segment can be used to predict the status of the pandemic for future dates.

\cite{mehrizi2020detection} have applied the mPRUTF algorithm to detect change points that have occurred in the transformed COVID-19 datasets from March 10, 2020 until April 30, 2021, for Australia, Canada, United Kingdom and the United States. We have applied our proposed post-detection inference approaches to these datasets. The $\% \, 95$ post-detection confidence intervals  for the identified change points are provided in Figure \ref{fig:pdi_covid}. For example, for Canada, based on the polyhedron and local post-detection inference approaches, the change points located on March 26, 2020; April 9, 2020; May 11, 2020; August 31, 2020 and January 12, 2021 are significant. For the United Kingdom, based on the polyhedron and local post-detection approaches, only the change point located on June 22, 2020 is insignificant. Moreover, the figure shows that the confidence intervals for March 18, 2021 in Canada and for February 23, 2021 in the United States derived from polyhedron post-detection method are very wide and skewed. These observations certify the results provided in Theorem \ref{thm:cilength.zttrunc.proj2}.


\begin{figure}[!t]
\begin{center}
\begin{subfigure}{.44\textwidth}
  \centering
  \includegraphics[width=1\linewidth]{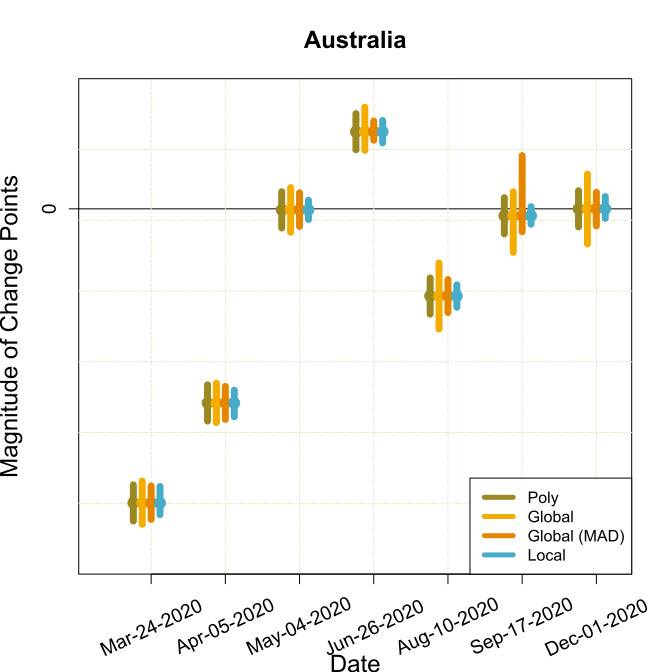}
\end{subfigure}
\begin{subfigure}{.44\textwidth}
  \centering
  \includegraphics[width=1\linewidth]{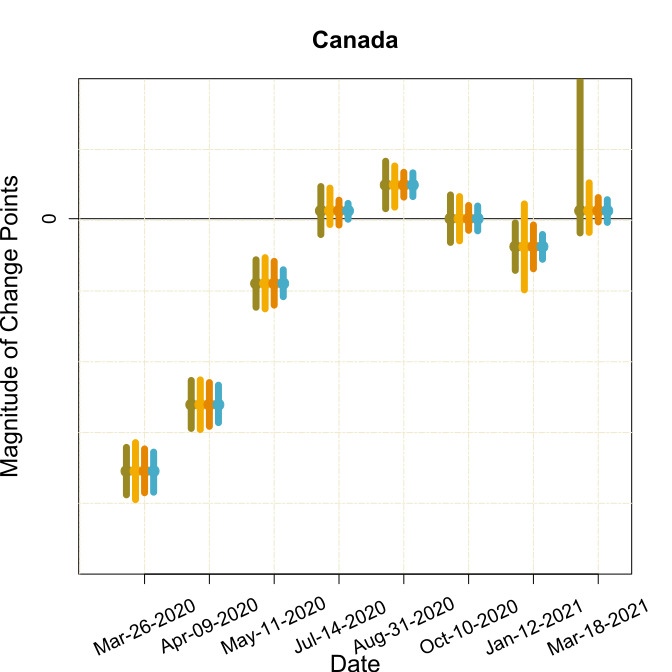}
\end{subfigure}
\\
\vspace{.5cm}
\begin{subfigure}{.44\textwidth}
  \centering
  \includegraphics[width=1\linewidth]{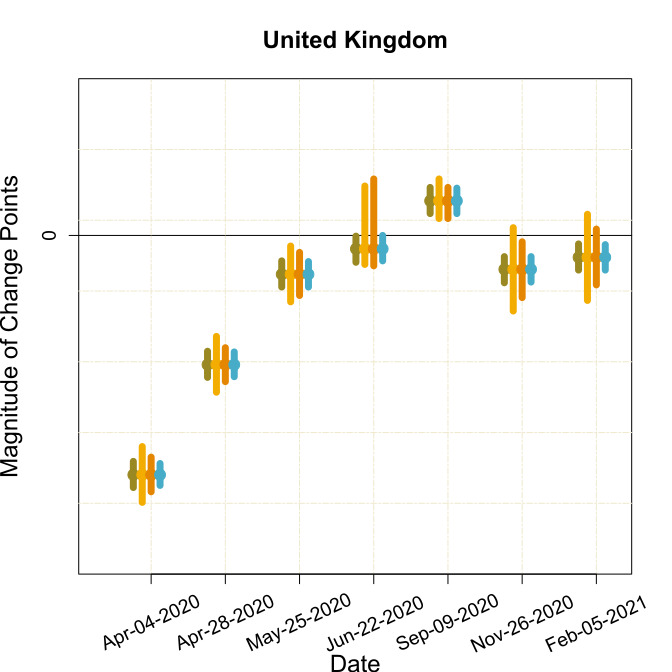}
\end{subfigure}
\begin{subfigure}{.44\textwidth}
  \centering
  \includegraphics[width=1\linewidth]{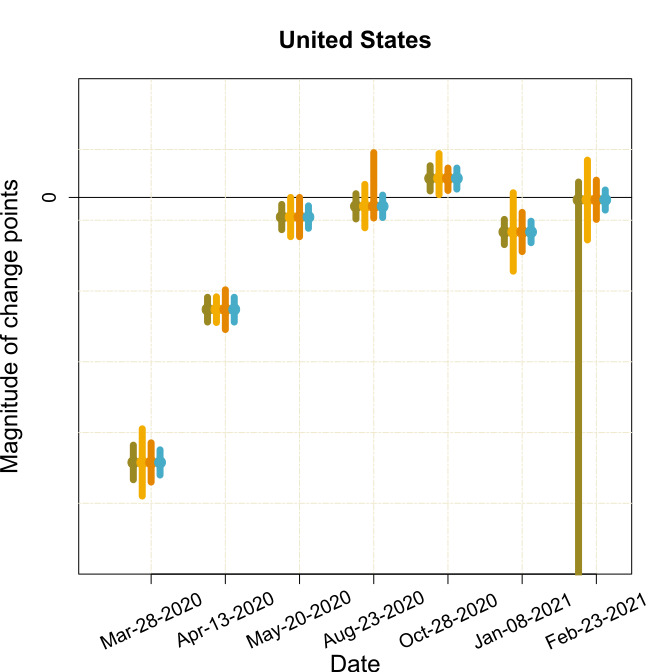}
\end{subfigure}

\caption [Post-Detection Confidence Intervals For COVID-19 Datasets] { Valid post-detection confidence intervals for COVID-19 datasets.. For each estimated change point, $\% \, 95$ post-detection confidence intervals obtained from polyhedron, global with both pooled and MAD variance estimations, and local post-detection inference approaches are displayed.}
\label{fig:pdi_covid}
\end{center}
\end{figure}



\end{example}

\section{Discussion}
\label{sec:discussion}
In this manuscript, we have attempted to quantify the uncertainty of change points estimated by the PRUTF algorithm, put forward by \cite{mehrizi2020detection}. By applying trend filtering, PRUTF yields an efficient algorithm to detect change points in a piecewise polynomial signal. We have provided a post-detection inference framework to compute valid $p$-values for the significance of the estimated change points. We have also constructed confidence intervals for the magnitude of such estimated change points. These inferences have been executed by implementing conditional inference through three distinct conditional detection events, hence, giving polyhedron, global and local post-detection inference procedures.  We have also shown that the global and local post-detection inferences lead to higher-powered tests by conditioning on much smaller detection events.

For future work in this area, there are several possible ideas. We would like to apply our proposed post-detection approaches to the extensively used change point detection algorithms such as binary segmentation, its variants and $\ell_0-$norm segmentation. Our focus in this work has been on the independent and identically distributed random noises to precisely highlight the main ideas. However, our proposed approaches are not confined only to these types of random noises. One possible extension would be to incorporate more complex random noises. For example, we can consider heavy-tailed distributions or even auto-correlated random noises. The extension of the proposed approaches in this work will allow us to provide inferential tools for a wide range of applications.

\appendix

\section*{Appendix A}
\renewcommand{\theequation}{A.\arabic{equation}}
\renewcommand{\thesubsection}{A.\arabic{subsection}}

\subsection{Proof of Theorem \ref{thm:ztrunc.proj2}}
\label{prf:ztrunc.proj2}
\begin{enumerate}[label=\alph*)]
    \item Note that
    \begin{align}\label{Ay.proj2}
       \mbf A\mbf y =\mbf A \big( \mbf P_{\eta}+\mbf P-\mbf P_{\eta} \big)\, \mbf y
       = \left(\frac{\mbf A\bsy\eta}{\|\bsy\eta\|}\right)\,\frac{\bsy\eta^T\mbf y}{\|\bsy\eta\|}+ \mbf A \mbf V,
    \end{align}
    which allows us to rewrite $\big\{ \mbf A \mbf y \geq \mbf q \big\}$ as
    \begin{align*}
       \Big\{ \mbf A \mbf y \geq \mbf q \Big\}= \Big\{\, \sigma\, \Big[\mbf A \bsy\eta/\|\bsy\eta\|\Big] Z+ \Big[ \mbf A \mbf V - \mbf q \Big] ~\geq~ 0 \Big\}.
    \end{align*}
    This representation leads to the truncation set
    \begin{align*}
       \big[ \mca V_Z^-\,,\, \mca V_Z^+ \big]= \bigcap_{i=1}^{ | \row{\mbf A} |}\Bigg\{z\in \mbb R:~ \sigma\, \Big[\mbf A \bsy\eta/\|\bsy\eta\|\Big]_i z+ \Big[ \mbf A \mbf V - \mbf q \Big]_i\geq 0 \Bigg\}.
    \end{align*}
    Observe that solving the inequality in the above set with respect to $z$ depends on the sign of $\rho_i=\big[\mbf A \bsy\eta/\|\bsy\eta\|\big]_i\,$. Therefore,
    \begin{align*}
       \Big\{ z \in \mbb R:~ \sigma \bsy\rho_i\, z+ \big[\mbf A \mbf V - \mbf q \big]_i \geq 0\,  ,\, \text{\footnotesize $i=1,\ldots, \big| \row{\mbf A} \big|$}  \Big\}= \left\{\begin{array}{ll}
        z \geq \frac{ \big[ \mbf q - \mbf A \mbf V \big]_i}{\sigma \bsy\rho_i}    &  i: ~ \bsy\rho_i>0,  \\[8pt]
        z \leq \frac{ \big[ \mbf q - \mbf A \mbf V \big]_i}{\sigma \bsy\rho_i}    &  i: ~ \bsy\rho_i<0,    \\[8pt]
        0 \leq \big[ {\scriptstyle \mbf A \mbf V - \mbf q } \big]_i   &  i: ~ \bsy\rho_i=0.
       \end{array}
       \right.
    \end{align*}
    The preceding statement leads to the interval $\big[ \mca V_Z^-\,,\, \mca V_Z^+ \big]$, provided $\mca V_Z^0>0$, where $\mca V_Z^-=\mca V_Z^- (\mbf V)$, $\mca V_Z^+=\mca V_Z^+(\mbf V)$ and $\mca V_Z^0=\mca V_Z^0 (\mbf V)$ are provided in \eqref{ztrunc.bounds.proj2}. Given $\mbf V$, these truncation boundaries are fixed. Therefore, the distribution of $Z$ given $\big\{ \mbf V,\, \mbf A \mbf y\geq \mbf q \big\} $ is equivalent to the distribution of a normal distribution constrained to the interval $\big[ \mca V_Z^-\,,\, \mca V_Z^+ \big]$. Hence
    \begin{align*}
        Z ~\Big|~ \big\{ \mbf V,\, \mbf A \mbf y\geq \mbf q \big\} \sim \trm{TN} \big(\bsy\eta^T\mbf f \, ,\, 1\, ,\, \big[ \, \mca V_Z^-\, ,\, \mca V_Z^+ \, \big]\, \big).
    \end{align*}
    
    
    \item Applying the probability integral transform for \eqref{ztrunc.law.proj2} yields 
    \begin{align*}
        1-\Phi^{[\mca V_Z^-,\mca V_Z^+]}\,(Z)\, \Big |\, \big\{\mbf V,\, \mbf A\mbf y\geq \mbf q \big\} ~\sim~ U \big( \,0 \, ,\, 1\, \big).
    \end{align*}
    For every $0\leq u \leq 1$, by marginalizing the above statement over $\mbf V$,  we have 
    \begin{align*}
        \Pr\bigg( 1- &\Phi^{[\mca V_Z^-,\mca V_Z^+]}\,(Z)  \leq u\, \Big |\, \mbf A\mbf y \geq \mbf q \bigg)
        \\[8pt]
        &=\int\displaylimits_{_{\mbf V}}  \Pr \left( 1-\Phi^{[\mca V_Z^-,\mca V_Z^+]}\,(Z)\leq u\, \Big |\, \mbf V\,,\, \mbf A\mbf y\geq \mbf q \right)\, \Pr \Big(\mbf V \, \big |\, \mbf A\mbf y\geq \mbf q \Big)\,\mbf d \mbf v 
        \\[8pt]
        &=u\,\int \displaylimits_{_{\mbf V}}  \Pr(\mbf V \,\Big |\, \mbf A\mbf y\geq \mbf q)\,\mbf d \mbf v\, = \, u
    \end{align*}
    which establishes Equation \eqref{tn.stat.proj2}.
\end{enumerate}

\section{Proof of Theorem \ref{thm:ttrunc.proj2}}
\label{prf:ttrunc.proj2}
\begin{enumerate}[label=\alph*)]
    \item Let $W= \big\| \big(\mbf I-\mbf P+\mbf P_{\eta} \big)\, \mbf y \big\|^2$, since both projection matrices $\mbf P$ and $\mbf P_{\eta}$ are symmetric and idempotent, then
    \begin{align*}
        W&= \big\|(\mbf I-\mbf P)\, \mbf y \big\|^2 +\left(\frac{\bsy\eta^T\mbf y}{\|\bsy\eta\|}\right)^2+ 2\, \bigg( \frac{\bsy\eta^T(\mbf I -\mbf P)\, \mbf y}{\|\bsy\eta\|}\bigg) \, \bigg(\frac{\bsy\eta^T\mbf y}{\|\bsy\eta\|} \bigg)
        \\[10pt]
        &=\, \widehat{\sigma}^2\, \bigg[ d+ T^2+ 2\, \bigg( \frac{\bsy\eta^T(\mbf I -\mbf P)\, \mbf y}{\widehat{\sigma} \|\bsy\eta\|} \bigg) T \bigg].
    \end{align*}
    From \eqref{Ay.proj2}, the polyhedron set can be rewritten as
    \begin{align*}
    \Big\{ \mbf A\mbf y ~\geq~ \mbf q \Big\}= \bigg\{ \Big(\frac{\mbf A\bsy\eta}{\|\bsy\eta\|}\Big)\,\frac{\bsy\eta^T\mbf y}{\|\bsy\eta\|}+ \mbf A \mbf V ~\geq~ \mbf q \bigg\}.
    \end{align*}
    Multiplying both sides of the above inequality by $ W/\, \widehat{\sigma}^2$ yields the truncation set
    {\small
    \begin{align}
    \big[\mca V_T^-\, , \, \mca V_T^+\big] &=\bigcap_{i=1}^{ |\row{\mbf A}|} \Bigg\{ t\in \mbb R :~ \Big( \frac{W \rho_i}{\widehat{\sigma}} \Big)\, t+ \Big[ \mbf A \mbf V - \mbf q \Big]_i\,\bigg( d+ t^2+ 2\, \frac{ \bsy\eta^T (\mbf I -\mbf P)\, \mbf y}{\widehat{\sigma} \|\bsy\eta\|} \, t \bigg) \geq 0 \Bigg\} 
    \\[10pt]
    &=\bigcap_{i=1}^{ |\row{\mbf A}|} \Bigg\{ t\in \mbb R :~ 
    \Big[ \mbf A \mbf V - \mbf q \Big]_i\, t^2 + \bigg( 2\, \frac{\bsy\eta^T(\mbf I -\mbf P)\, \mbf y}{\widehat{\sigma} \|\bsy\eta\|}\, \Big[ \mbf A \mbf V - \mbf q \Big]_i\, +\, \frac{W \rho_i}{\widehat{\sigma}} \bigg)\, t 
    \\[8pt]
    &\hspace{2.85cm} +\, \Big[ \mbf A \mbf V - \mbf q \Big]_i\, d  \geq 0 \Bigg\}.
    \end{align}}
    In the above intersection, for each $i, ~ i= 1\, ,\, \ldots\, ,\, |\row{\mbf A}|$, the inequality can be solved explicitly. Note that $\big[\mca V_T^-\, , \, \mca V_T^+\big]$ is a bounded interval (see Theorem \ref{thm:ciboundary.zttrunc.proj2}). 
    
    On the other hand, $W$ can also be decomposed as
    \begin{align}
        W= \big\| \big( \mbf I-\mbf P+\mbf P_{\eta} \big)\, \mbf y \big\|^2=\, \big\|\mbf y \big\|^2\, -\, \big\| \big( \mbf P-\mbf P_{\eta} \big)\, \mbf y \big\|^2.
    \end{align}
     Therefore, conditioning on $\big\{ \mbf V\, ,\, \|\mbf y\|^2\big\}$ is equivalent to conditioning on $\big\{ \mbf V, \, W\big\}$. This means that, given $\big\{ \mbf V\, ,\, \|\mbf y\|^2\big\}$, the truncation boundaries $\mca V_T^-=\mca V_T^- \big(\mbf V,\, W \big)$ and $\mca V_T^+=\mca V_T^+ \big( \mbf V,\, W \big)$ are fixed. Therefore, the distribution of $T$ given $\big\{ \mbf V\, ,\, \|\mbf y\|^2\big\}$ is equivalent to the distribution of a $t$ distribution with $d$ degrees of freedom, constrained to $\big[\mca V_T^-\, , \, \mca V_T^+\big]$. Hence,
    \begin{align*}
        T~ \Big|~\Big\{\, \mbf  V,\, \big\| \mbf y \big\|^2\, ,\, \mbf A \mbf y \geq \mbf q \Big\} ~\sim~ {\rm Tt} \big( \bsy\eta^T \mbf f ,\, 1,\, d,\, [\mca V_T^-,\mca V_T^+] \big)
    \end{align*}
    
    \item Similar to part ($b$) of Theorem \ref{thm:ztrunc.proj2}, by marginalization of \eqref{ttrunc.law.proj2} over $\big\{ \mbf V\, ,\, W \big\}$, we obtain the statement in \eqref{tt.stat.proj2} 
\end{enumerate}

\section{Proof of Theorem \ref{thm:cilength.zttrunc.proj2}}
\label{prf:cilength.zttrunc.proj2}

\begin{enumerate}[label=(\alph*)]
    \item The result of this part has been proved in \cite{kivaranovic2020length}. Here, we give the proof in our notation. Lemma A.4 of the latter reference shows that for $b_k<\infty$,
    \begin{align*}
        \lim_{\mathit{z}\longrightarrow b_k^-}\, \big( b_k -z \big)\, L_Z(z) & ~=~ -\sigma^2 \log \big( 1-\alpha/2 \big),
        \\[8pt]
        \lim_{\mathit{z}\longrightarrow b_k^-}\, \big( b_k- z \big)\, U_Z(z) & ~=~ -\sigma^2 \log \big(\alpha/2 \big).
    \end{align*}
    The above equations together leads to
    \begin{align*}
        \lim_{z \longrightarrow b_k^-}\, \big( b_k-z \big) \big[\, U_Z(z)-L_Z(z)\, \big]= \sigma^2 \log\, \big((2-\alpha)/\alpha \big)
    \end{align*}
     This limit states that there exits an  $\epsilon>0$, such that 
     \begin{align*}
         U_Z(z)-L_Z(z) ~\geq~ \frac{\sigma^2\, \log\, \big( (2-\alpha)/\alpha \big)}{2\, \big( b_k-z \big)} 
         \qquad\qquad\trm{for any }~ z \in \big( b_k-\epsilon\, , \, b_k \big)\, \medcap\, \mca S_Z.
     \end{align*}
     Now, let $f^*= \inf \Big\{ f_{\mu,\,\sigma^2}^{\,\mca S_Z}(z):~ z \in \big( b_k- \epsilon\, ,\, b_k \big)\, \medcap\, \mca S_Z \Big\}$. Clearly,  $f^*>0$ as $f_{\mu,\, \sigma^2}^{\,\mca S_Z}$ is a probability density function. Therefore,
     \begin{align}\label{z.ci.infty.proj2}
         E \Big[ U_Z(Z) - L_Z(Z)\, \Big|\, Z\in \mca S_Z \Big] & ~=~ \int_{\, z\, \in\, \mca S_Z}\, \big[ U_Z(z)- L_Z(z) \big]\,  f_{\mu,\, \sigma^2}^{\,\mca S_Z}(z)\,  dz
         \\[10pt]
         & \geq~ \frac{ \sigma^2 \log \big( (2-\alpha)/ \alpha \big) }{2} \int_{z\, \in\, ( b_k- \epsilon\, ,\, b_k )\, \bigcap\, \mca S_Z} \frac{f_{\mu,\,\sigma^2}^{\,\mca S_Z}(z)}{b_k-z}\, dz
         \\[10pt]
         & \geq\, \frac{ \sigma^2 \log \big( (2-\alpha)/ \alpha \big)\, f^* }{2} \int_{z\, \in\, ( b_k- \epsilon\, ,\, b_k )\, \bigcap\, \mca S_Z}\, \frac{1}{b_k-z}\, dz=\infty.
     \end{align}
    This can be similarly shown for $a_1> -\infty$.
    
    \item From the result in part (a), it suffices to show that
    \begin{align*}
        E \Big[ U_T(T)-L_T(T) \Big] ~\geq~ E \Big[ U_Z(Z)- L_Z(Z) \Big].
    \end{align*}
    To this end, we show that a random variable $T\sim {\rm Tt}\, \big(\mu\, ,\, \sigma^2\, ,\, d\, ,\, \mca S_T \big)$ has a heavier tail than a random variable $Z \sim {\rm TN}\, \big( \mu\, ,\, \sigma^2\, ,\, \mca S_T \big)$. It is known that a random variable $Y$ has heavier tails than $X$  if and only if
    \begin{align*}
        \lim_{t\longrightarrow \infty}\frac{1-F_Y(t)}{1-F_X(t)}=\infty.
    \end{align*}
    This equation is equivalent to
    \begin{align*}
        \lim_{t\longrightarrow \infty}\frac{f_Y(t)}{f_X(t)}=\infty.
    \end{align*}
    Now, let $f_{\mu,\,\sigma^2}^{\,\mca S_T}(\cdot)$ and $g_{\mu,\,\sigma^2,\, d}^{\,\mca S_T}(\cdot)$, denote the density function of $Z$ and T, respectively. Therefore,
    \begin{align*}
        \lim_{x\longrightarrow \infty} \frac{g_{\mu,\,\sigma^2,\, d}^{\,\mca S_T}(x)}{f_{\mu,\,\sigma^2}^{\,\mca S_T}(x)}=& \lim_{y \longrightarrow \infty} \frac{g_{ d}^{\,\mca S_T} (y) }{ f_{0,\, 1}^{\,\mca S_T}(y)}= \lim_{y \longrightarrow \infty} \exp\left\{\log g_{ d}^{\,\mca S_T}(y)- \log f_{0,\, 1}^{\,\mca S_T}(y) \right\}
        \\[8pt]
        =& \exp\left\{ \lim_{y \longrightarrow \infty}- \frac{d+1}{2}\, \log \left( 1+ \frac{y^2}{d} \right)+ \frac{1}{2}\, y^2+C\right\}
        \\[8pt]
        =& \exp\left\{ \lim_{y \longrightarrow \infty} y^2 \left[ -\frac{d+1}{2\, y^2} \log\, \left( 1+\frac{y^2}{d} \right)+ \frac{1}{2} \right]+C\right\},
    \end{align*}
    where in the first equality we use the change variable $y=(x-\mu)/\sigma$. Since
    \begin{align*}
        \lim_{y \longrightarrow \infty}\, \frac{1}{ y^2}\, \log\, \left( 1+\frac{y^2}{d} \right) =0,
    \end{align*}
    we have
    \begin{align*}
        \lim_{x\longrightarrow \infty} \frac{g_{\mu,\,\sigma^2,\, d}^{\,\mca S_T}(x)}{f_{\mu,\,\sigma^2}^{\,\mca S_T}(x)}= \infty.
    \end{align*}
    The preceding result proves that the truncated $t$ distribution has heavier tails than truncated normal.


    Now, since $\big[ L_Z(Z)\, ,\, U_Z(Z)]$ and $\big[ L_T(T)\, ,\, U_T(T)]$ are confidence intervals at the same level for truncated normal and truncated $t$ distributions, respectively, and $U_Z(Z)-L_Z(Z)$ and $ U_T(T)-L_T(T)$ are two non-negative random variables such that
    \begin{align*}
        U_T \big( T \big)\, -\, L_T \big( T \big) ~ \overset{a.s.}{\geq} ~  U_Z \big( Z \big)\, -\, L_Z \big( Z \big) \,.
    \end{align*}
    Hence,
    \begin{align*}
        E \Big[\, U_T(T)\, -\, L_T(T)\, \Big]& ~ \geq ~ E \Big[\, U_Z(Z)\, -\, L_Z(Z)\, \Big]\, .
    \end{align*}
    The above equation and Equation \eqref{z.ci.infty.proj2} together lead to the statement in \eqref{ttrunc.infinite.expect.length}.
\end{enumerate}


\section{Proof of Theorem \ref{thm:ciboundary.zttrunc.proj2}}
\label{prf:ciboundary.zttrunc.proj2}
\begin{enumerate}[label=(\alph*)]
    \item From Equation \eqref{ztrunc.bounds.proj2}, it can be viewed that $\mca V_Z^-$ and $\mca V_Z^+$ are unbounded if both sets $\big\{i:\, \rho_i< 0\big\}$ and $\big\{i:\, \rho_i >0 \big\}$ are empty, where $\rho_i=\big[ \mbf A\bsy\eta/\|\bsy\eta\| \big]_i$, for $i=1\, ,\, \ldots\, ,\, \big| \row{\mbf A} \big|$. This condition implies $\rho_i=0$ for all $i$ which in turn requires that $\mbf A\bsy\eta=\mathbf{0}$ for all $\bsy\eta\neq \mathbf{0}$. Hence every vector $\bsy\eta\neq \mathbf{0}$ is orthogonal to the row space of the matrix $\mbf A$, which contradicts with the fact that the conditioning set is a polyhedron.
    \item From \eqref{ttrunc.bounds.proj2}, $\mca V_T^-$ and $\mca V_T^+$ are unbounded if the inequality
    \begin{align}\label{ttrunc.quadform}
        \Big[ \mbf A \mbf V - \mbf q \Big]_i\, t^2 + \bigg( 2\, \frac{\bsy\eta^T(\mbf I -\mbf P)\, \mbf y}{\widehat{\sigma} \|\bsy\eta\|}\, \Big[ \mbf A \mbf V - \mbf q \Big]_i + \frac{W \rho_i}{\widehat{\sigma}} \bigg)\, t +\, \Big[ \mbf A \mbf V - \mbf q \Big]_i\, d  \geq 0
    \end{align}
    always holds for any $i=1,\ldots, \big| \row{\mbf A} \big|$. This requires that the quadratic equation in \eqref{ttrunc.quadform} to have at most one real root, i.e.,
    \begin{align*}
    \Bigg( 2\, \frac{\bsy\eta^T(\mbf I -\mbf P)\, \mbf y}{\widehat{\sigma} \|\bsy\eta\|}\, \Big[ \mbf A \mbf V - \mbf q \Big]_i\, +\, \frac{W \rho_i}{\widehat{\sigma}} \Bigg)^2 -\, 4\, d\, \Big[ \mbf A \mbf V - \mbf q \Big]_i^2 \leq 0 \,, 
    \end{align*}
    as well as the coefficient sign of $t^2$ must be positive, i.e., $\big[ \mbf A \mbf V - \mbf q \big]_i > 0$. According to \eqref{ztrunc.bounds.proj2}, this latter condition occurs if and only if $\rho_i=0$, for all $i=1,\ldots, \big| \row{\mbf A} \big|$, which is impossible as shown in part (a).
    
\end{enumerate}

\section{Proof of Theorem \ref{thm:global.proj2}}
\label{prf:global.proj2}

\begin{enumerate}[label=(\alph*)]
    \item Since $s_{\widehat{\tau}_j}$ is the sign of $\big[ \mbf D \widehat{\mbf f}\, \big]_{\widehat{\tau}_j} = \mbf D  _{\widehat{\tau}_j} \widehat{\mbf f}$, hence, $ s_{\widehat{\tau}_j}\, \mbf D  _{\widehat{\tau}_j} \widehat{\mbf f} \geq 0$. From the primal-dual relationship in \eqref{primal.dual.proj2}, we get
    \begin{align}\label{sDf.positive.proj2}
        s_{\widehat{\tau}_j}\, \mbf D_{\widehat{\tau}_j} \big(\mbf y-\mbf D^T \widehat{\mbf u}_{\lambda_j} \big) \geq 0,
    \end{align}
    where $\lambda_j$ is the value of the regularization parameter associated with $\widehat{\tau}_j$.  Now, the goal is to find a range for $Z_j^{\trm{glo}}$ for which $\widehat{\tau}_j$ is a change point. Recall that from KKT conditions of the dual problem \eqref{tf.dual.obj.proj2}, if $\widehat{\tau}_j$ is a change point then $\widehat{ u}_{\lambda_j,\, \widehat{\tau}_j}= s_{\widehat{\tau}_j}\, \lambda_j$. Applying the decomposition $\mbf y= \mbf P_{\eta}\, \mbf y + \big(\mbf I - \mbf P_{\bsy\eta} \big)\, \mbf y$ into \eqref{sDf.positive.proj2} and setting $\widehat{ u}_{\lambda_j,\, \widehat{\tau}_j}= s_{\widehat{\tau}_j}\, \lambda_j$, we have 
    \begin{align*}
        s_{\widehat{\tau}_j}\, \mbf D_{\widehat{\tau}_j} \left(\frac{\bsy\eta}{\|\bsy\eta \|^2}\, \bsy\eta^T\, \mbf y + \big(\mbf I - \mbf P_{\bsy\eta} \big)\, \mbf y - s_{\widehat{\tau}_j}\, \lambda_j\, \mbf D_{\widehat{\tau}_j}^T - \mbf D_{-\widehat{\tau}_j}^T\, \widehat{\mbf u}_{\lambda_j,\, -\widehat{\tau}_j}\right).
    \end{align*}
    Hence, setting $\bsy\eta= \mbf D_{\widehat{\tau}_j}^T$ in the preceding equation, for $s_{\widehat{\tau}_j}=1$, 
    \begin{align}
        \mbf D_{\widehat{\tau}_j} \mbf y\, \geq\, \lambda_j\, \mbf D_{\widehat{\tau}_j}\, \mbf D_{\widehat{\tau}_j}^T + \mbf D_{\widehat{\tau}_j}\, \mbf D_{-\widehat{\tau}_j}^T\, \widehat{\mbf u}_{\lambda_j,\, -\widehat{\tau}_j},
    \end{align}
    and, for $s_{\widehat{\tau}_j}=-1$, 
    \begin{align}
        \mbf D_{\widehat{\tau}_j} \mbf y\, \leq\, -\lambda_j\, \mbf D_{\widehat{\tau}_j}\, \mbf D_{\widehat{\tau}_j}^T + \mbf D_{\widehat{\tau}_j}\, \mbf D_{-\widehat{\tau}_j}^T\, \widehat{\mbf u}_{\lambda_j,\, -\widehat{\tau}_j}.
    \end{align}
    Therefore, given $\mbf D_{\widehat{\tau}_j}$, $\lambda_j$, $\widehat{\mbf u}_{\lambda_j}$, and $\sigma^2$, the quantity $Z_j^{\,\trm{glo}}= \frac{\mbf D_{\widehat{\tau}_j}\, \mbf y}{\sigma\, \|\mbf D_{\widehat{\tau}_j}^T\|}$ is restricted  to regions $\big(-\infty\, ,\, \mca V_{Z_j}^{\,-(glo)}\big]$ and $\big[\mca V_{Z_j}^{\,+(glo)}\, ,\, \infty)$ where $\mca V_{Z_j}^{-(glo)}$ and $\mca V_{Z_j}^{+(glo)}$ are given in \eqref{glo.zbounds.proj2}.

    \item In the same fashion as part (a), given $\mbf D_{\widehat{\tau}_j}$, $\lambda_j$ and $\widehat{\mbf u}_{\lambda_j}$, the quantity $T_j^{\,\trm{glo}}=\frac{\mbf D_{\widehat{\tau}_j} \mbf y}{\widehat{\sigma}_j^{(\,\trm{glo})}\, \|\mbf D_{\widehat{\tau}_j}^T\|}$ is restricted  to the regions $\big(-\infty\, ,\, \mca V_{T_j}^{\,-(glo)}\big]$ and $\big[\mca V_{T_j}^{\,+(glo)}\, ,\, \infty \big)$ where $\mca V_{T_j}^{-(glo)}$ and $\mca V_{T_j}^{+(glo)}$ are given in \eqref{glo.tbounds.proj2}.
\end{enumerate}

\section{Proof of Theorem \ref{thm:upperbound.zttrunc.global.proj2}}
\label{prf:upperbound.ttrunc.proj2}
We will prove the inequality for the truncated $t$ distribution. A similar proof can be applied to the truncated normal distribution which is also provided in \cite{kivaranovic2020length}. Suppose $T \sim t\, \big( \mu\, ,\, \sigma^2\, ,\, d)$ is a location-scale $t$ distribution with cumulative distribution function $G_{\mu,\, \sigma^2,\, d}\, (\cdot)$.
For any $0<\gamma<1$, define $Q_\gamma(t)$ via
\begin{align*}
    G_{Q_{\gamma},\, \sigma^2,\, d} \big( t \big)= \trm{Pr}_{Q_{\gamma},\, \sigma^2,\, d} \big( T\leq t \big)= \gamma,
\end{align*}
or equivalently, $Q_\gamma(t)=t- \sigma\,G_{d}^{-1} (\gamma)$. Thus, the cumulative distribution function of $T$ truncated to $\mca S=\bigcup_{\,i=1}^{\,m} \big( c_i\,,\,d_i\big)$, is upper bounded by
\begin{align}\label{gd.right.proj2}
    G_{Q_{\gamma},\, \sigma^2,\, d}^{\,\mca S} \big( t \big)&=\, \frac{\trm{Pr}_{{Q_{\gamma},\, \sigma^2,\, d}} \big( T\leq t\,\, \medcap \,\, T\in \mca S \big)}{\trm{Pr}_{{Q_{\gamma},\, \sigma^2,\, d}} \big(  T\in \mca S \big)}
    \\[8pt]
    &\leq\, \frac{\Pr_{Q_{\gamma},\, \sigma^2,\, d} \left(T\leq t\right)}{\Pr_{Q_{\gamma},\, \sigma^2,\, d}\left(T\in \mca S \right)}\, \leq\, \frac{\gamma}{\underline{g}},
\end{align}
where $ \underline{g}= \inf_{\, _{\mu}}\,\, \trm{Pr}_{\mu\, ,\,\sigma^2\, ,\, d}\, \big( T\in \mca S \big)$. On the other hand, from the inequality $\Pr(A\medcap B)\geq \Pr(A)+ \Pr(B)-1$, we notice that
\begin{align}\label{gd.left.proj2}
    G_{Q_{\gamma},\, \sigma^2,\, d}^{\,\mca S} \big( t \big)&=\, \frac{\trm{Pr}_{{Q_{\gamma},\, \sigma^2,\, d}} \big( T\leq t\,\, \medcap \,\, T\in \mca S \big)}{\trm{Pr}_{{Q_{\gamma},\, \sigma^2,\, d}} \big(  T\in \mca S \big)}
    \\[8pt]
    &\geq \frac{\trm{Pr}_{{Q_{\gamma},\, \sigma^2,\, d}} \big( T\leq t \big)\, +\, \trm{Pr}_{{Q_{\gamma},\, \sigma^2,\, d}} \big(T \in \mca S \big)\, -1 }{\trm{Pr}_{{Q_{\gamma},\, \sigma^2,\, d}} \big(  T\in \mca S \big)}
    \\[8pt]
    &\geq\, \inf_\mu\,\, \frac{\trm{Pr}_{{Q_{\gamma},\, \sigma^2,\, d}} \big( T\leq t \big)\, +\, \trm{Pr}_{{\mu,\, \sigma^2,\, d}} \big(T \in \mca S \big)\, -1 }{\trm{Pr}_{{\mu,\, \sigma^2,\, d}} \big(  T\in \mca S \big)}
    \\[8pt]
    &= \frac{\gamma+ \overline{g} -1}{\overline{g}},
\end{align}
where $\overline{g}= \sup_{\, \mu}\,\, \trm{Pr}_{\mu\, ,\,\sigma^2\, ,\, d}\, \big( T\in \mca S \big)$.
Equations \eqref{ttrunc.ci.proj2} and \eqref{gd.right.proj2} imply that
\begin{align*}
    G_{Q_{\frac{\alpha}{2}\, \underline{g}} ,\, \sigma^2,\, d}^{\,\mca S} \big( t \big)\, \leq\, \frac{\alpha}{2}\, =\, G_{U(T),\, \sigma^2,\, d}^{\,\mca S} \big( t \big).
\end{align*}
Because $ G_{\mu,\, \sigma^2,\, d}^{\,\mca S} \big( t \big)$ is a decreasing function with respect to $\mu$, hence,
\begin{align*}
    U(T)& ~\overset{a.s.}{\leq}~ Q_{\frac{\alpha}{2}\, \underline{g}}(T)= T- \sigma\, G_d^{-1} \Big( \frac{\alpha}{2}\, \underline{g} \Big)\,.
\end{align*}
Similarly, from  \eqref{ttrunc.ci.proj2} and \eqref{gd.left.proj2}, we obtain
\begin{align*}
    ~~L(T)& ~\overset{a.s.}{\geq}~ Q_{1-\frac{\alpha}{2}\, \overline{g}}(T)= T- \sigma\, G_d^{-1} \Big( 1- \frac{\alpha}{2}\, \overline{g} \Big)\,.
\end{align*}
Therefore, from the last two equations, we have
\begin{align}\label{first.upper.bound.proj2}
    U(T)-L(T) &~\overset{a.s.}{\leq}~ \sigma \bigg[\, G_d^{-1} \Big( 1- \frac{\alpha}{2}\, \overline{g} \Big) - G_d^{-1} \Big( \frac{\alpha}{2}\, \underline{g} \Big) \bigg]
    \\[8pt]
    &~=~ \sigma \bigg[\, G_d^{-1} \Big( 1- \frac{\alpha}{2}\, \overline{g} \Big) + G_d^{-1} \Big( 1-\frac{\alpha}{2}\, \underline{g} \Big) \bigg]
    \\[8pt]
    &~\leq~  2\, \sigma\, G_d^{-1} \Big( 1- \frac{\alpha}{2}\, \underline{g} \Big)\,.
\end{align}
Note that the last equality in the preceding statement is obtained from $1- \frac{\alpha}{2}\, \overline{g} \leq 1- \frac{\alpha}{2}\, \underline{g}$. Now, consider the minimization problem
\begin{align*}
    \inf_\mu\,\, \trm{Pr}_{{\mu,\, \sigma^2,\, d}} \Big( T< d_1 ~~ \trm{or} ~~ T>c_m \Big),
\end{align*}
which minimizes at $\frac{d_1 + c_m}{2}$, with the minimum value $g_0= 2\, G_d \Big(-\frac{c_m- d_1 }{ 2\, \sigma} \Big)$. Observe that $g_0 \leq \underline{g}$, and since $G^{-1}\, (\cdot)$ is an increasing function, then
\begin{align*}
    G_d^{-1} \Big( 1- \frac{\alpha}{2}\, \underline{g} \Big) ~\leq~ G_d^{-1} \Big( 1- \frac{\alpha}{2}\, g_0 \Big).
\end{align*}
Lastly, to obtain the final upper bound in \eqref{t.upper.bound.proj2}, it suffices to show 
\begin{align}\label{second.inequlity.upper.bound.proj2}
    G_d^{-1} \Big( 1-\frac{\alpha}{2}\, g_0 \Big) ~\leq~  G_d^{-1} \Big( 1-\frac{\alpha}{2}\, \Big)\, +\, \frac{c_m- d_1}{2\, \sigma} \,,
\end{align}
or equivalently,
\begin{align}\label{equivalent.second.inequlity.upper.bound.proj2}
    G_d^{-1} \Big(\frac{\alpha}{2}\, g_0 \Big) ~\geq~  G_d^{-1} \Big( \frac{\alpha}{2} \Big)\, -\frac{c_m- d_1}{2\, \sigma} \,,
\end{align}

\noindent
Given the condition \eqref{t.upper.bound.condition.proj2}, we note that,
\begin{align}\label{t.condition.cd.proj2}
    \alpha\,\, G_d \Big( -\frac{c_m- d_1}{2\, \sigma} \Big) \,\geq\, G_d \Big( G_d^{-1} \big( \alpha/2 \big)- \frac{c_m- d_1}{2\, \sigma} \Big), \qquad\qquad \trm{for} \quad d \geq 3.
\end{align}
Hence, by plugging in the value of $g_0$, we have
\begin{align*}
    \frac{\alpha}{2}\, g_0= \alpha\,\, G_d \left( -\frac{c_m- d_1 }{ 2\, \sigma}\right) \,\geq\, G_d \left( G_d^{-1} \left( \frac{\alpha}{2} \right)- \frac{ c_m- d_1 }{ 2\, \sigma} \right).
\end{align*}
Applying the increasing function $G_d^{-1} (\cdot)$ to both sides of the above inequality leads to \eqref{equivalent.second.inequlity.upper.bound.proj2}, and in turn \eqref{second.inequlity.upper.bound.proj2}. Now, Equations \eqref{first.upper.bound.proj2} and \eqref{second.inequlity.upper.bound.proj2} together yield 
\begin{align*}
    U(T)- L(T) ~\overset{a.s.}{\leq}~  2\, \sigma\, G_d^{-1} \Big( 1-\frac{\alpha}{2} \Big)\, +\, \big( c_m- d_1 \big).
\end{align*}
Finally, the upper bound associated with the truncated $t$ distribution in \eqref{t.upper.bound.proj2} will be achieved by replacing $L(T)$, $U(T)$, $\sigma$, $d_1$, $c_m$ and $d$ with $L_{T_j}^{\trm{glo}}$, $U_{T_j}^{\trm{glo}}$, $\widehat{\sigma}_j^{\, (\trm{glo})}$, $d^{(\trm{glo})}$, $\mca V_{T_j}^{-(\trm{glo})}$, $\mca V_{T_j}^{+(\trm{glo})}$ and $d^{\,\mathrm{(glo)}}$, respectively.

\begin{figure}[!ht]
\begin{center}
\begin{subfigure}{.45\textwidth}
  \centering
  \includegraphics[width=1\linewidth]{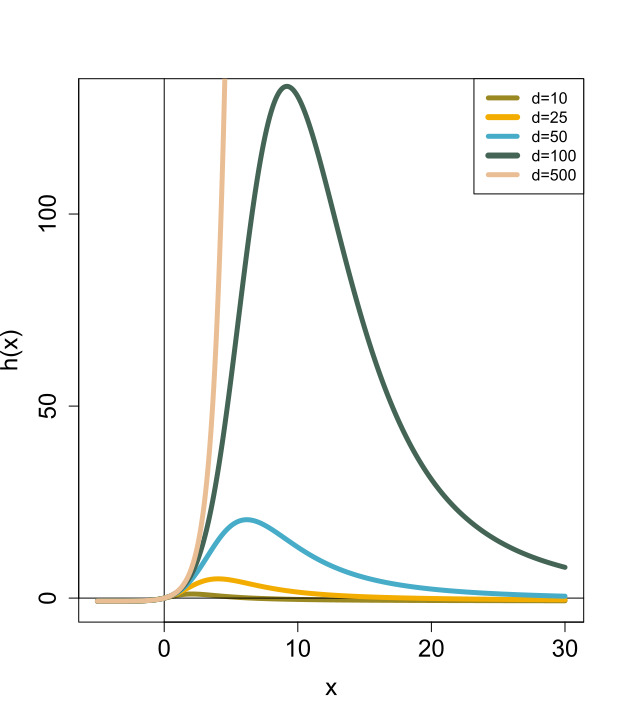}
  \caption{Function $h(x)$.}
\end{subfigure}
\quad
\begin{subfigure}{.45\textwidth}
  \centering
  \includegraphics[width=1\linewidth]{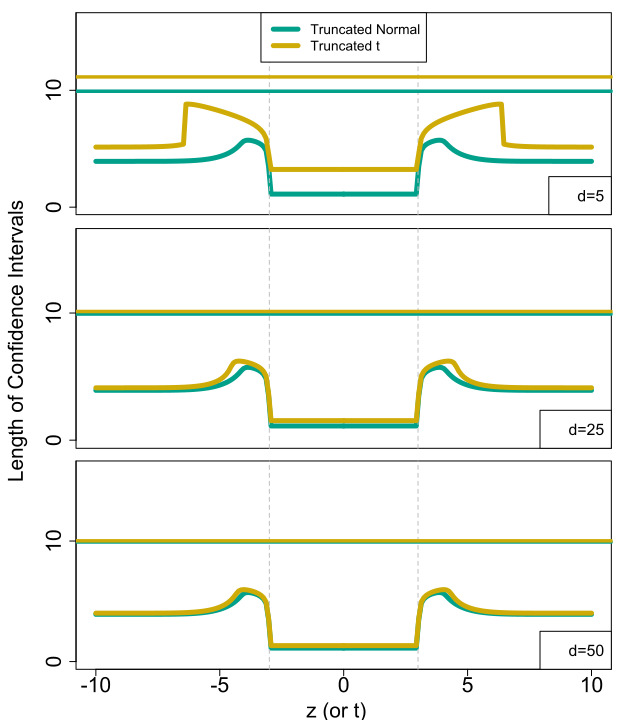}
  \caption{Truncation set $\mca S= (-\infty\, ,\, -3) \cup (3\, ,\, \infty)$.}
\end{subfigure}

\caption[Structure of $h(x)$ And the Upper Bounds For Confidence Intervals Using Global And Local Post-Detection Inferences]{ Panel (a) displays the function $h(x)$ which, for moderate values of $d$, is a unimodal function with two roots, zero and a positive value $x_0$. Also, $h(x) \geq 0$ for  $x \in [0\, ,\, x_0]$. Panel (b) shows the lengths of confidence intervals for a truncated normal distribution and a truncated $t$ distribution, under various values of $d$. The solid horizontal lines in panel (b) shows the upper bounds given in Theorem \ref{thm:upperbound.zttrunc.global.proj2}, for both distributions.}
\label{fig:upperbound.proj2}
\end{center}
\end{figure}

\noindent
To explore the condition \eqref{t.condition.cd.proj2},  define
\begin{align*}
    h(x)=\frac{\alpha\, G_d (-x)}{G_d\left( G_d^{-1}(\alpha/2)- x\right) }-1\, , \qquad\qquad \trm{for} \quad d \geq 3\, .
\end{align*}
The function $h(x)$ has two roots, one at zero and one at a positive value $x_0$. Moreover, $h(x) > 0$, for $x \in \big( 0\, ,\, x_0 \big)$.
Notice that the positive root $x_0$ diverges as degrees of freedom $d$ increases. More specifically, for large $d$, the truncated $t$ distribution behaves similar to normal one, and hence $h(x)$ becomes an increasing function for all $x > 0$. See Figure \ref{fig:upperbound.proj2}, panel (a). Therefore, for large$d$, the statement in \eqref{t.condition.cd.proj2} always holds. For small to moderate values of $d$, \eqref{t.condition.cd.proj2} holds under the condition $ \frac{c_m- d_1}{2\, \sigma} \, \in \, \big( 0\, ,\, x_0 \big]$. We have visualized the behaviour of lengths of confidence intervals and their upper bounds in Figure \ref{fig:upperbound.proj2}, panel (b), for various values of $d$.


\renewcommand\refname{Bibliography}
\bibliographystyle{apalike}
\bibliography{Project2}

\end{document}